\documentclass[a4paper,11pt]{article}
\usepackage{amssymb}
\usepackage{amsfonts}
\usepackage{graphicx}
\usepackage{epstopdf}
\usepackage{dcolumn}
\usepackage{amsmath}
\usepackage{latexsym,bm}
\usepackage{geometry}

\def \be {\begin{equation}}
\def \ee {\end{equation}}
\def \bea {\begin{eqnarray}}
\def \eea {\end{eqnarray}}
\def \nn {\nonumber}

\def \a {\alpha}
\def \b {\beta}

\def \d {\delta}

\def \m {\mu}
\def \n {\nu}
\def \k {\kappa}

\def \s {\sigma}
\def \r {\rho}
\def \o {\omega}

\def \th {\theta}
\def \Th {\Theta}

\def \t {\tau}
\def \dag {\dagger}
\def \p {\partial}

\def\bd{\begin{document}}
\def\ed{\end{document}}
\def\nn{\nonumber}
\def\bea{\begin{eqnarray}}
\def\eea{\end{eqnarray}}
\let\bm=\bibitem
\let\la=\label

\def\N{{\cal N}}
\def\sst{\scriptscriptstyle}
\def\thetabar{\bar\theta}
\def\Tr{{\rm Tr}}
\def\one{\mbox{1 \kern-.59em {\rm l}}}

\def\a{\alpha}      \def\da{{\dot\alpha}}
\def\b{\beta}       \def\db{{\dot\beta}}
\def\c{\gamma}  \def\C{\Gamma}  \def\cdt{\dot\gamma}
\def\d{\delta}  \def\D{\Delta}  \def\ddt{\dot\delta}
\def\e{\epsilon}        \def\vare{\varepsilon}
\def\f{\phi}    \def\F{\Phi}    \def\vvf{\f}
\def\h{\eta}
\def\k{\kappa}
\def\l{\lambda} \def\L{\Lambda}
\def\m{\mu} \def\n{\nu}
\def\o{\omega}
\def\P{\Pi}
\def\r{\rho}
\def\s{\sigma}  \def\S{\Sigma}
\def\t{\tau}
\def\th{\theta} \def\Th{\Theta} \def\vth{\vartheta}
\def\X{\Xeta}
\def\z{\zeta}
\def\w{\wedge}
\def\u{\underline}
\def\hs{\hspace}

\def\cA{{\cal A}} \def\cB{{\cal B}} \def\cC{{\cal C}}
\def\cD{{\cal D}} \def\cE{{\cal E}} \def\cF{{\cal F}}
\def\cG{{\cal G}} \def\cH{{\cal H}} \def\cI{{\cal I}}
\def\cJ{{\cal J}} \def\cK{{\cal K}} \def\cL{{\cal L}}
\def\cM{{\cal M}} \def\cN{{\cal N}} \def\cO{{\cal O}}
\def\cP{{\cal P}} \def\cQ{{\cal Q}} \def\cR{{\cal R}}
\def\cS{{\cal S}} \def\cT{{\cal T}} \def\cU{{\cal U}}
\def\cV{{\cal V}} \def\cW{{\cal W}} \def\cX{{\cal X}}
\def\cY{{\cal Y}} \def\cZ{{\cal Z}}

\def\ua{\underline{\alpha}} \def\ubb{\underline{\beta}}
\def\ug{\underline{\gamma}}
\def\ub{\underline{\phantom{\alpha}}\!\!\!\beta}
\def\uc{\underline{\phantom{\alpha}}\!\!\!\gamma}
\def\um{\underline{\mu}} \def\un{\underline{\nu}}
\def\ud{\underline\delta}
\def\ue{\underline\epsilon}
\def\una{\underline a}\def\unA{\underline A}
\def\unb{\underline b}\def\unB{\underline B}
\def\unc{\underline c}\def\unC{\underline C}
\def\und{\underline d}\def\unD{\underline D}
\def\une{\underline e}\def\unE{\underline E}
\def\unf{\underline{\phantom{e}}\!\!\!\! f}\def\unF{\underline F}
\def\unm{\underline m}\def\unM{\underline M}
\def\unn{\underline n}\def\unN{\underline N}
\def\unp{\underline{\phantom{a}}\!\!\! p}\def\unP{\underline P}
\def\unq{\underline{\phantom{a}}\!\!\! q}
\def\unQ{\underline{\phantom{A}}\!\!\!\! Q}
\def\unH{\underline{H}}
\def\ul{\underline}

\def\As {{A \hspace{-6.4pt} \slash}\;}
\def\bs {{b \hspace{-6.4pt} \slash}\;}
\def\Ds {{D \hspace{-6.4pt} \slash}\;}
\def\ds {{\del \hspace{-6.4pt} \slash}\;}
\def\ss {{\s \hspace{-6.4pt} \slash}\;}
\def\ks {{ k \hspace{-6.4pt} \slash}\;}
\def\ps {{p \hspace{-6.4pt} \slash}\;}
\def\pas {{{p_1} \hspace{-6.4pt} \slash}\;}
\def\pbs {{{p_2} \hspace{-6.4pt} \slash}\;}

\def\Fh{\hat{F}}
\def\Vh{\hat{V}}
\def\Xh{\hat{X}}
\def\ah{\hat{a}}
\def\xh{\hat{x}}
\def\yh{\hat{y}}
\def\ph{\hat{p}}
\def\xih{\hat{\xi}}

\def\psit{\tilde{\psi}}
\def\Psit{\tilde{\Psi}}
\def\tht{\tilde{\th}}

\def\At{\tilde{A}}
\def\Qt{\tilde{Q}}
\def\Rt{\tilde{R}}
\def\Nt{\tilde{N}}

\def\at{\tilde{a}}
\def\st{\tilde{s}}
\def\ft{\tilde{f}}
\def\pt{\tilde{p}}
\def\qt{\tilde{q}}
\def\vt{\tilde{v}}
\def\nt{\tilde{n}}

\def\delb{\bar{\partial}}
\def\bz{\bar{z}}
\def\bD{\bar{D}}
\def\bB{\bar{B}}

\def\bk{{\bf k}}
\def\bl{{\bf l}}
\def\bp{{\bf p}}
\def\bq{{\bf q}}
\def\br{{\bf r}}
\def\bx{{\bf x}}
\def\by{{\bf y}}
\def\bR{{\bf R}}
\def\bV{{\bf V}}

\def\d{\delta}\def\D{\Delta}\def\ddt{\dot\delta}

\def\p{\partial} \def\del{\partial}
\def\xx{\times}
\def\uno{\mbox{1 \kern-.59em {\rm l}}}

\def\trp{^{\top}}
\def\inv{^{-1}}
\def\dag{{^{\dagger}}}

\def\pr{\prime}
\def\rar{\rightarrow}
\def\lar{\leftarrow}
\def\lrar{\leftrightarrow}

\title{Phase Structure of Higher Spin Black Hole}
\author{
Bin Chen$^{1,2,3}$\footnote{bchen01@pku.edu.cn},\,
Jiang Long$^{1,2}$\footnote{lj301@pku.edu.cn}\,
and
Yi-Nan Wang$^{1}$\footnote{ynwang@pku.edu.cn}
}
\date{}

\begin{document}

\maketitle

\begin{center}
{\it
$^{1}$Department of Physics, Peking University, Beijing 100871, P.R. China\\
\vspace{2mm}
$^{2}$State Key Laboratory of Nuclear Physics and Technology, Peking University, Beijing 100871, P.R. China\\
\vspace{2mm}
$^{3}$Center for High Energy Physics, Peking University, Beijing 100871, P.R. China\\
}
\vspace{10mm}
\end{center}

\date{}
\begin{abstract}

In this paper, we investigate the phase structures of the black holes with one single higher spin hair, focusing specifically on the spin 3 and spin $\tilde 4$ black holes. Based on dimensional analysis and the requirement of having consistent thermodynamics,
we derive an universal formula relating the entropy and the conserved charges for arbitrary $AdS_3$ higher spin black holes. Then we use it to study the phase structure of the higher spin black holes. We find that there are six branches of solutions in the spin 3 gravity, eight branches of solutions in the spin $\tilde{4}$ gravity and twelve branches of solutions in the $G_2$ gravity. In each case, all branches are related by a simple angle shift in the entropy functions. In the spin 3 case, we reproduce all the results found before. In the spin $\tilde{4}$ case, we find that in the low temperature it is at the $BTZ$ branch while in the high temperature it transits to one of two other branches, depending on the signature of the chemical potential, a reflection of charge conjugate asymmetry found before. 
 \end{abstract}
 \newpage

 \section{Introduction}

 Black hole physics has been one of central topics in quantum gravity.
 One of the most important achievements in the 1970s is the discovery of black holes thermodynamics. Especially the area law of black hole entropy indicates the holographic principle in quantum gravity. As a concrete realization of the holographic principle, AdS/CFT correspondence was proposed by studying the brane physics. In the past ten or more years, black hole thermodynamics has obtained a revival after being embedded into AdS/CFT correspondence. It was found that the black hole configuration in AdS spacetime is  dual to the conformal field theory at finite temperature. The phase structure of the field theory could be understood from bulk black hole thermodynamics. This is the underlying physics in the recent applications of AdS/CFT correspondence to the study of QCD and other strongly coupled systems in condensed matter physics.

 On the other hand, the AdS/CFT correspondence has been expected to shed light on the nature of spacetime and gravity. One interesting investigation is the so-called high spin/CFT correspondence. The earlier study seemed to indicate a correspondence between AdS
 high spin gravity\cite{Vasiliev:1987,Vasiliev:1990,Vasiliev:1992}  and Yang-Mills theory in the free field limit\cite{Haggi-Mani:2000, Konstein:2000,Sundborg:2001,Mikhailov:2002}.
 More careful studies on the spectrum led to the conjecture of $HS/O(N)$ correspondence\cite{Klebanov:2002}\footnote{See \cite{Petkou},\cite{Sezgin} for a fermionic verison and \cite{ABJ} for the generalization to Chern-Simons vector theory.}. The perturbative check of the conjecture in $AdS_4$ has been push forward in the past few years by the computation of three-point functions in \cite{Giombi I:2010, Giombi II:2010}. However, the non-perturbative aspects are not clear except some exact solutions  \cite{Sezgin:2005,Iazeolla:2007,Didenko:2009,Iazeolla:2011,Iazeolla:2012}, whose physical interpretations are unknown right now. Fortunately, in the past few years, $AdS_3$ higher spin theory has opened a new window to study various aspects of HS/CFT correspondence. It  turns out to be as fruitful as $AdS_4$ higher spin gravity. Some important developments include the asymptotical symmetry analysis\cite{Henneax:2010, Theisen:2010}, the  Gaberdiel-Gopakumar conjecture\cite{Gaberdiel:2011} and the identification of higher spin black holes in $AdS_3$\cite{Per kraus:2011}. Among them, the high spin black hole is of particular interest.

 The most remarkable feature of higher spin black hole is that it changes the conventional notions on spacetime and black holes\cite{Per kraus:2011,Ammon:2011}. As in a higher spin gravity, the diffeomorphism of graviton is modified by the higher spin degrees of freedom, the conventional methods in dealing with the black hole and its thermodynamics make no much sense. One must use gauge invariant quantities to study the black hole\cite{Per kraus:2011}. Moreover, the spin 3 black hole in AdS$_3$ gives an explicit example which shows the power of holography. Without the holographic descriptions as the guideline, it is not clear  how to interpret the solutions found in the higher spin $AdS_3$ gravity so far. After imposing appropriate requirements on the holonomy, which is gauge invariant, and with the help of dual CFT, the thermodynamics of the higher spin black holes could be well-defined. Moreover, the higher spin black hole has been studied from HS/CFT point of view as well, and consistent picture has emerged\cite{kraus:2011,Matthias:2012}. For a nice review and complete references, see \cite{HSBH}. Very recently, the higher spin black holes with just one single higher spin hair have been classified\cite{truncated,Chen:2012cz}. These black holes are based on the truncated higher spin gravity and are easier to study in many aspects. Similarly, the thermodynamics of these black holes have been defined consistently.

 Since the higher spin black holes are consistent with the first law of thermodynamics, it is imminent to explore their phase structures. Here one needs to know the free energy (or partition function) of the higher spin black holes. In \cite{Theisen:2012} this problem has been tackled systematically at first time. However, the variables used there do not have direct $CFT$ interpretations hence the formulas obtained can not be used to explore thermodynamics directly. In \cite{David:2012}, the author revisit this problem in the spin 3 gravity in the spirit of \cite{Theisen:2012}. The most interesting part of their work is the discovery of multiple branches of solutions in the spin 3 higher spin gravity. It was found that  the $BTZ$ branch only dominates in the low temperature and it must transit to another branch at a critical temperature.

Here we try to explore the phase structures of the higher spin black holes more systematically. At first we derive an entropy formula (and a partition function formula) by using simple dimensional analysis and the first law of thermodynamics. This formula should valid for arbitrary higher spin $AdS_3$ black holes. The most remarkable fact is that the formula is actually independent of the holonomy equations\footnote{Actually, this fact was shown explicitly by angular quantization in \cite{Theisen:2012}.}. Then we use this formula to solve the holonomy equations in the spin 3 and the spin $\tilde{4}$ gravity. We find that there are multiple branches from the holonomy equations, if we do not require that the entropy reduces to the one of BTZ black hole in the limit of zero higher spin charge. We find that there are six branches for the spin 3 black hole, eight branches for the spin $\tilde{4}$ black hole and twelve branches for the $G_2$ black hole. We work out the exact entropy function in each branch, from which we are allowed to investigate the phase structure in more details.  For the spin 3 black hole, we reproduce all the results found in \cite{David:2012}. This gives a consistent check of our treatment. For the spin $\tilde{4}$ black hole, we find that all the branches are charge conjugate asymmetric, as was first discovered in \cite{truncated}. When we restrict ourselves in the positive entropy and positive temperature region, there are only three branches (Branch 1, (5,+,-) and (8,+,+)) left. At low temperature, though the entropy of Branch 1, which is usually called BTZ branch as it could reduce to BTZ black hole in the limit of zero spin 4 charge,  is smaller than the ones of other two branches, only Branch 1 is stable in the sense that the specific heat is non-negative only in it. Therefore Branch 1 is thermodynamically favored in the low temperature region. In the high temperature region, Branch 1 disappears, the system transits to Branch (5,+,-) or Branch (8,+,+), depending on the signature of the chemical potential. Moreover, we notice that there are  two sharp windows in which the system is thermodynamically unstable.

The structure of this paper is as follows. In section 2 we derive the formula of the entropy and the partition function by dimensional analysis and thermodynamics. In section 3 we explore the phase structure of the spin 3 black hole. In section 4 we explore the phase structure of the spin $\tilde{4}$ black hole.  We end this paper by some conclusion and discussions.

\section{General Setup}

In this section, we derive an universal entropy formula for arbitrary AdS$_3$ higher spin gravity. To be concrete, we focus on the $SL(N,R)$ higher spin gravity and use the principal embedding. However, the derivation can be simply generalized to $Sp(2N,R),SO(2N+1,R)$ and $G_2$ gravity which was investigated recently in \cite{truncated}. Some earlier efforts on this problem were  presented in \cite{Theisen:2012}.

Let us consider a higher spin black hole with higher spin charges $\mathcal{L}_n, \bar{\mathcal{L}}_n$ where $n=2,3,\cdots,N$. The conjugate chemical potentials are labeled by $\alpha_n, \bar{\alpha}_n$. For $n=2$, $\mathcal{L}_2$ is the stress tensor in the boundary CFT and $\alpha_2$ is the inverse temperature $\tau$. We denote the partition function of this higher spin black hole as $Z$ and the entropy as $S$. The partition function $Z$ is\footnote{The coefficient $4\pi^2 i$ is dependent of the normalization of the higher spin charge and the chemical potential. In the discussion of the spin $\tilde{4}$ and $G_2$ gravity below, we will change the convention to match the results given in \cite{truncated}.}
\be
Z=Tr \exp{i 4\pi^2(\sum_{n=2}^{N}\alpha_n\mathcal{L}_n-\sum_{n=2}^{N}\bar{\alpha}_n\bar{\mathcal{L}}_n)}
\ee
where $n$ is summed up from 2 to N. By taking the logarithmic of the partition function we find
\be
\ln Z=S+4\pi^2 i(\sum_{n=2}^{N}\alpha_n\mathcal{L}_n-\sum_{n=2}^{N}\bar{\alpha}_n\bar{\mathcal{L}}_n).
\ee
To satisfy the first law of thermodynamics we require
\be
\mathcal{L}_n=-\frac{i}{4\pi^2}\frac{\partial\ln Z}{\partial\alpha_n},
\ee
or equivalently,
\be
\alpha_n=\frac{i}{4\pi^2}\frac{\partial S}{\partial\mathcal{L}_n}.\label{che}
\ee

Now we consider the problem: what is the most general entropy formula which is consistent with the previous results? To be concrete, we assume all the higher spin charges are positive. Firstly, note that the entropy is a function of the higher spin charges $\mathcal{L}_n$. Secondly, from the dimensional analysis we learn that the entropy should have the form\footnote{We are a bit sloppy here since there are other dimensional constants $G$ and $l$ in the theory. This form of the entropy can be checked in the following sections.}
\be
S(\mathcal{L}_n)=\sum_{\{a_n\}} c_{a_2\cdots a_N} \prod_{n=2}^{N}\mathcal{L}_n^{a_n/n},\la{en}
\ee
where the summation is over all the possible partition of $\{a_n\}$ with the constrains that
\be
\sum_{n=2}^{N}a_n=1.
\ee
 The coefficient $c_{a_2\cdots a_N}$ can be arbitrary. Therefore the entropy satisfies the equation that
\be
S=\sum_{n=2}^N n\mathcal{L}_n\frac{\partial S}{\partial \mathcal{L}_n}.
\ee
Then we use the relation (\ref{che}) to find that
\be
S=-4\pi^2 i\sum_{n=2}^N n\alpha_n\mathcal{L}_n.
\ee
After including the bar term, we find an universal formula of the entropy for the  higher spin black hole
\be
S=-4\pi^2 i\sum_{n=2}^N n\alpha_n\mathcal{L}_n+4\pi^2 i\sum_{n=2}^N n\bar{\alpha}_n\bar{\mathcal{L}}_n. \label{entropy}
\ee
And  the partition function should be
\be
\ln Z=-4\pi^2 i\sum_{n=2}^N (n-1)\alpha_n\mathcal{L}_n+4\pi^2 i\sum_{n=2}^N (n-1)\bar{\alpha}_n\bar{\mathcal{L}}_n.\label{partition}
\ee

As a first check of our formula, we use it to match the results given in \cite{David:2012}. For the spin 3 higher spin gravity, we choose the principal embedding, and consider the non-rotating case which is defined as
\be
\tau=-\bar{\tau},\hs{3ex}\mathcal{L}=\bar{\mathcal{L}}, \hs{3ex} \mathcal{W}=-\bar{\mathcal{W}},\hs{3ex}\alpha=\bar{\alpha}.
\ee
We also need the relations between $\tau, \alpha$ and the temperature $T=\frac{1}{\beta}$  the potential $\mu$ in the non-rotating case
\be
\tau=\frac{i\beta}{2\pi}, \hs{3ex} \alpha=-\tau\mu.
\ee
Then we find that the entropy for the spin 3 higher spin black hole satisfy the relation
\be
S=\frac{1}{T}(8\pi\mathcal{L}-12\pi\mathcal{W}\mu) \label{entropyrelation3}
\ee
   which is exactly the one given in \cite{David:2012}. In the same way, we can determine the partition function as
\be
\ln Z=\frac{1}{T}(4\pi\mathcal{L}-8\pi\mathcal{W}\mu)
\ee
which has also been given in \cite{David:2012} up to a sign. The sign difference comes from the definition of their grand potential. Note that in their paper, the above formulas were obtained by a sophisticated subtraction of the on-shell action and using the holonomy equations. However, according to our analysis, we find that the forms of the entropy and the partition function are the results of the consistency of the thermodynamics and the dimensional analysis, and have nothing to to with the holonomy equations. Nevertheless, the agreement with the results in \cite{David:2012} gives a support of the formulas (\ref{entropy},\ref{partition}) and prove the consistency of our formula with the holonomy equations in the spin 3 gravity. However, our results are more general and can be used to arbitrary higher spin gravity. As we will show in Sect. 4, they could be applied to the study of the phase of the spin $\tilde 4$ black hole.

Another remarkable fact is that our formula could also be valid for non-principal embedding\footnote{Here we only give the observation. It is important to have the strict derivation.}. As a check, we still use the results of the spin 3 black hole appearing in \cite{David:2012}. In the spin 3 black hole case, the spin 3 black hole flows to a UV fixed point which is depicted by the diagonal embedding and the corresponding UV CFT has a $W_3^{(2)}$ symmetry. The UV CFT is deformed by a spin 3/2 relevant operator and  flows to IR fixed point. The corresponding charges are
the stress tensor $\hat{\mathcal{L}}$ and the spin 3/2 charge $\mathcal{G}$. Their corresponding chemical potentials are denoted as $\tau$ and $\hat{\alpha}$, with the potential being $\lambda=-\frac{\hat{\alpha}}{\tau}$.  Then from the result we obtained previously, the entropy and the partition function should be
\be
\hat{S}=\frac{1}{T}(8\pi\hat{\mathcal{L}}-6\pi\lambda\mathcal{G}),
\ee
\be
\ln \hat{Z}=\frac{1}{T}(4\pi\hat{\mathcal{L}}-2\pi\lambda\mathcal{G}).
\ee
These results are in perfect match with those in \cite{David:2012}.

\section{Phase Structure in Spin 3 Black Hole}

In this section, we revisit the phase structure of the spin 3 black hole, from a different point of view. Actually, we show that there are multiple branches from the holonomy equations, without imposing boundary condition. We determine the entropy function of each branch, which allows us to analyze the phase structure carefully.

Let us first define a dimensionless parameter $z=\sqrt{\frac{27k}{16\pi}}\frac{\mathcal{W}}{\mathcal{L}^{\frac{3}{2}}}$. Here we use a parameter different from the one used in \cite{Per kraus:2011} and subsequent studies. The reason is that the definition of $y=z^2$ in that paper hides the information of charge conjugation $\mathcal{W}\to-\mathcal{W}$. The entropy may be sensitive to the charge conjugation as first shown in \cite{truncated} in the spin ${\tilde 4}$ and $G_2$ cases. And from the work in \cite{David:2012}, it is obvious that for some branches, the entropy could be sensitive to the signature of the higher spin charge, even in the spin 3 case.

The entropy function can be still written in the form\footnote{Here we only include the  contribution of left-moving part. Similar treatment will be applied for the spin $\tilde{4}$ gravity case below.}
\be
S=4\pi\sqrt{2\pi k\mathcal{L}}f[z].
\ee
Note that the appearance of $\sqrt{\mathcal{L}}$ comes from dimensional analysis and the coefficients are just chosen to simplify the holonomy equations and they can be absorbed into the definition of the function $f[z]$.
Then the holonomy equations becomes
\bea
f[z]^2 - 9 (-2 + z^2)f'[z]^2=1,\\
-z f[z]^3 + 9 (-2 + z^2) f[z]^2f'[z] - 27 z (-2 + z^2) f[z]f'[z]^2 + 27 (-2 + z^2)^2f'[z]^3=0.
\eea
In this case, we do not impose the boundary condition that the entropy should be $BTZ$ entropy when the higher spin charge vanishes. This will lead to multiple branches of the solution. From the first holonomy equation, we find that the most general solution could be
\be
f[z]=\cos\th[z]
\ee
with
\be
\th[z]=\frac{1}{3}(\arcsin{\frac{z}{\sqrt{2}}}+a)
\ee
where $a$ is an arbitrary constant. We substitute this equation to the second holonmy equation and find
\be
\sin a=0
\ee
with $a$ being only fixed to
\be
a=n\pi, n=0,\pm1,\pm2,\cdots .
\ee
However, for the integer number $n$, only $n=0,1,2,3,4,5$ gives us different entropies. Then we find six possible forms of the entropy function
\bea
S_{1}&=&4\pi\sqrt{2\pi k\mathcal{L}}\cos\frac{1}{3}(\arcsin{\frac{z}{\sqrt{2}}}),\\
S_{2}&=&4\pi\sqrt{2\pi k\mathcal{L}}\cos\frac{1}{3}(\arcsin{\frac{z}{\sqrt{2}}}+\pi),\\
S_{3}&=&4\pi\sqrt{2\pi k\mathcal{L}}\cos\frac{1}{3}(\arcsin{\frac{z}{\sqrt{2}}}+2\pi),\\
S_4&=&4\pi\sqrt{2\pi k\mathcal{L}}\cos\frac{1}{3}(\arcsin{\frac{z}{\sqrt{2}}}+3\pi),\\
S_5&=&4\pi\sqrt{2\pi k\mathcal{L}}\cos\frac{1}{3}(\arcsin{\frac{z}{\sqrt{2}}}+4\pi),\\
S_6&=&4\pi\sqrt{2\pi k\mathcal{L}}\cos\frac{1}{3}(\arcsin{\frac{z}{\sqrt{2}}}+5\pi)
\eea
where the parameter $-\sqrt{2}\le z\le\sqrt{2}$ and the arcsin function takes value in the range $[-\frac{\pi}{2},\frac{\pi}{2}]$.
There are some remarkable features of these solutions:
\begin{itemize}
\item We do not use the non-rotating condition in the derivation. Hence the result is also valid for the rotating black holes. Since our result includes only left-moving part,  a general spin 3 higher spin black hole has $(6\times6=)$36 branches in all.
\item The existence of multiple branches is related to the periodicity of sinusoidal function.  It is easy to see that the entropy function of six branches can be connected to each other.
\item The extreme black hole is always at $z=\pm\sqrt{2}$, independent of the branches.
\item All the branches are related by a simple angle shift. Hence, according to different values of the angle appearing in the entropy function, we denote the branches as Branch 1, Branch 2, $\cdots$,Branch 6. Note that our definition of the branch is different from those in \cite{David:2012} where they used the root of spin 2 charge in the zero temperature limit to define the branches. Nevertheless, it turns out that our classification includes their results as a subset.
\end{itemize}

Let us have a closer look at these solutions. For simplicity, we consider the non-rotating case. The rotating case is also interesting, but we will not include them here.
\begin{enumerate}
\item We plot the $S/S_{0}-z$ diagram in Fig. \ref{S-zSpin3}, where $S_0$ is the entropy of $BTZ$ black hole.  There are six curves which corresponds to six branches. We use color and thickness to distinguish them in this section. From Branch 1 to Branch 6, they are labeled respectively by the curves in
\be
\mbox{(Red,Thin),(Green,Thin),(Blue,Thin),(Red,Thick),(Green,Thick),(Blue,Thick)}.\nn
\ee
The same color means that the two branches are related by a sign flipping on the entropy. Branch 4, Branch 5 and Branch 6 are related to Branch 1, Branch 2 and Branch 3 by a sign flipping on the entropy function. Let us determine the signature of the entropy in each branch. This can be easily determined from the entropy function. The result is that
\be
(1,+),\ (2,+),\ (3,-),\ (4,-),\ (5,-),\ (6,+)
\ee
The first number in the bracket denotes the branch, the second  denotes the signature of the entropy. It is clear that only Branch 1,2,6 have positive entropy. Hence, the physically allowed branch are Branch 1,2,6 according to the criteria that the entropy should be non-negative.
\begin{figure}
\centering
\includegraphics[height=9cm,width=12cm]{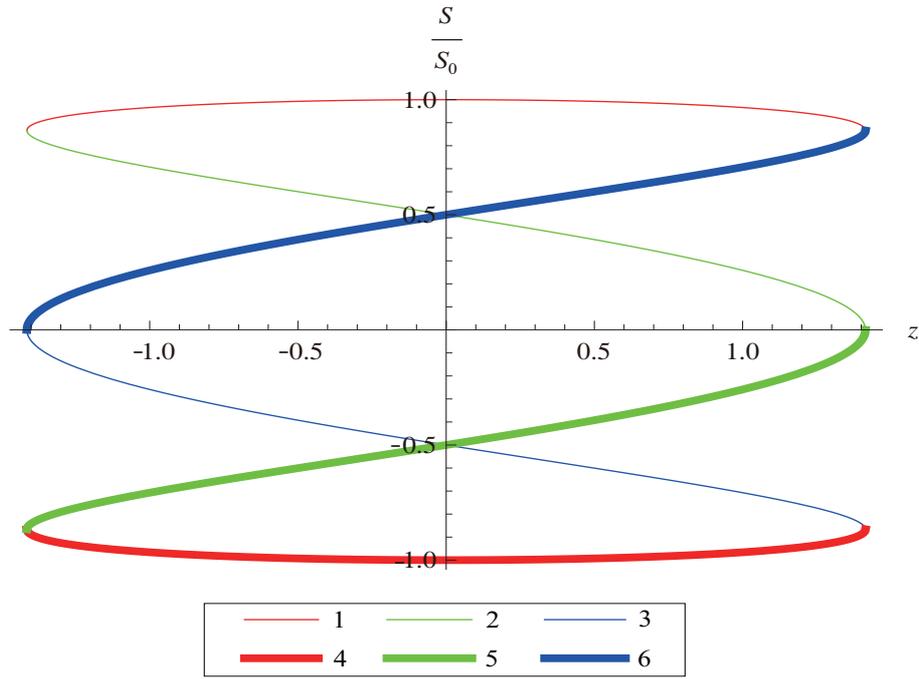}
\caption{$S/S_{0}-z$ relation: there are six branches, distinguished by color and thickness.}\label{S-zSpin3}
\end{figure}

\item Let us consider the charge conjugation. This can be found by $z\to-z$. We find that Branch 1 and Branch 4 are self conjugate invariant, and Branch 2(3) and Branch 6(5) are conjugated to each other. 

\item We plot $\mu T-z$ diagram of the six branches in Fig. \ref{T-zSpin3}. 
\begin{figure}
\centering
\includegraphics[height=9cm]{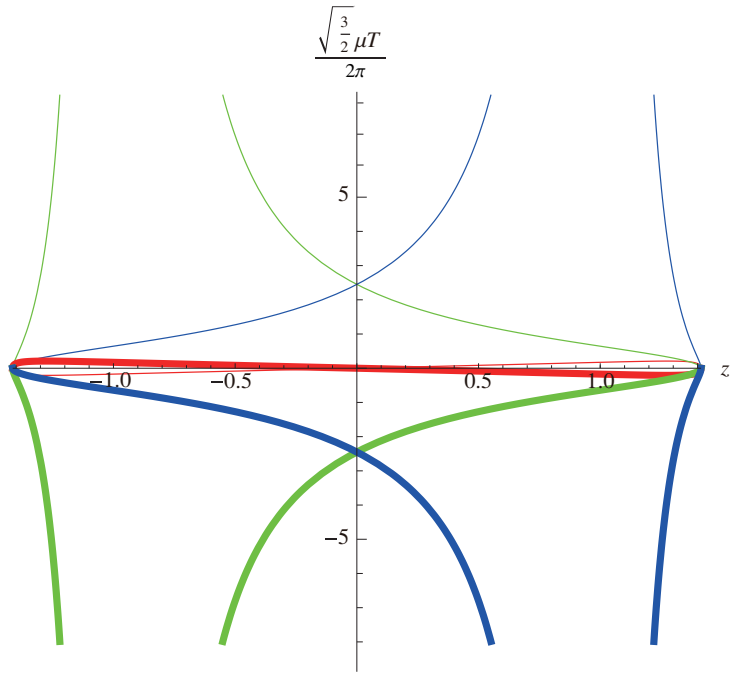}
\caption{$\mu T-z$ relation}\label{T-zSpin3}
\end{figure}
 We find that when the spin-3 charge is vanishing, namely, $z=0$, only Branch 1 and Branch 4 has a vanishing chemical potential. For other branches, they can have non-vanishing chemical potentials even when $z=0$. The values of $\mu T$ at $z=0$ in each branch are respectively
\be
(1,0),\ (2,\frac{3}{2\pi}),\ (3,\frac{3}{2\pi}),\ (4,0),\ (5,-\frac{3}{2\pi}),\ (6,-\frac{3}{2\pi})
\ee
Again, the first number in the bracket labels the branch, and the second number is the value of $\mu T$ at $z=0$. We find that the value of $\mu T$ in Branch 2 matches exactly the one in  Branch III discussed in \cite{David:2012}.

 We can see that Branch 1 and Branch 4 are restricted to a finite $\mu T$ range, so that they cannot exist at arbitrary high temperature. Since Branch 4 has a negative entropy,  we only consider Branch 1. The maximum value of $\mu T$ can be determined by the function
\be
\mu T=\frac{1}{2\pi}\sqrt{\frac{3}{2}}\zeta(z)=\frac{1}{2\pi}\sqrt{\frac{3}{2}}\frac{\sin\th(z)}{\sqrt{2-z^2}(\cos\th(z)+\frac{z\sin\th(z)}{\sqrt{2-z^2}})^2}.\label{mut}
\ee
For Branch 1, we can easily determine the extreme value of $\mu T$, which is at
\be
z=\frac{7-3\sqrt{3}}{\sqrt{2}},\ \mbox{or}\ z=-\frac{7-3\sqrt{3}}{\sqrt{2}}.
\ee
The corresponding $\mu T$ is
\be
\mu T=\frac{3}{16\pi}\sqrt{2\sqrt{3}-3},\ \mbox{or}\ \mu T=-\frac{3}{16\pi}\sqrt{2\sqrt{3}-3}
\ee
where the signature is determined by the signature of $\mu$. In \cite{David:2012}, the authors have chosen $\mu>0$, so they found only one point $\mu T$, which is in exact agreement with our result.

We can also read some other properties from Fig. \ref{T-zSpin3}. In Branch 2,3,5 and 6, the temperature can extend to $\infty$. The points where the temperatures go to $\infty$ are respectively
\be
(1,\phi),\ (2,-1),\ (3,1),\ (4,\phi),\ (5,-1),\ (6,1).
\ee
The first number denotes the branch and the second denotes the point that temperature tends to $\infty$. $\phi$ means that there is no point such that the temperature is $\infty$. However, we are a little sloppy here. Since what we plot is the $\mu T-z$ diagram, the information on the temperature is obscured.  To find out this information, we need to plot the $T/\sqrt{\mathcal{L}}-z$ diagram.

\item The $T/\sqrt{\mathcal{L}}-z$ diagram is plotted in Fig. \ref{L-zSpin3}.
\begin{figure}
\centering
\includegraphics[height=9cm]{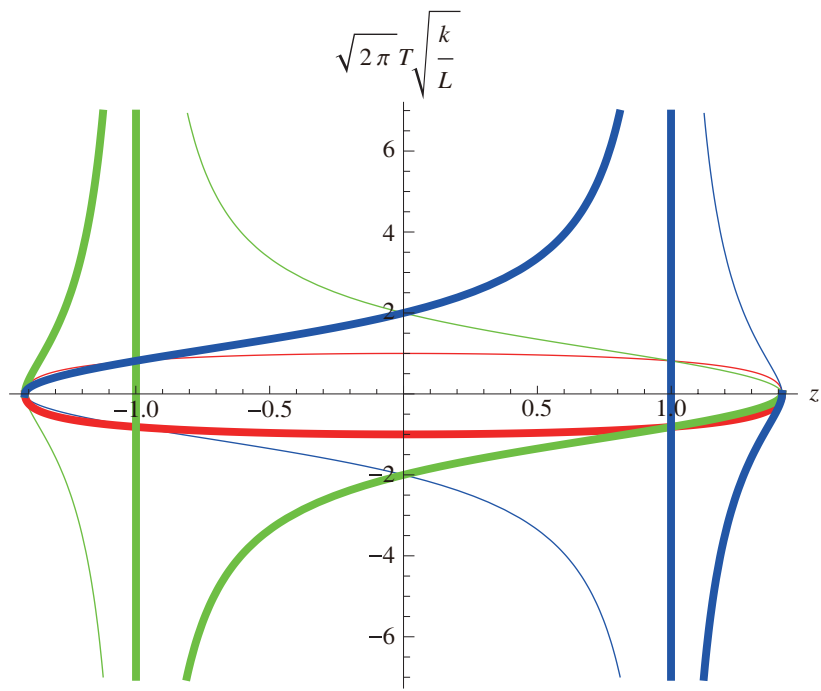}
\caption{$T/\sqrt{\mathcal{L}}-z$ relation}\label{L-zSpin3}
\end{figure}
We can summarize the information on the temperature read from the graph into the following table.

\begin{center}
\begin{tabular}{|c|c|c|}
\hline
Branch&z&T\\\hline
1&$-\sqrt{2},\  \sqrt{2}$&$0^+,\  \  0^+$\\\hline
2&$-\sqrt{2},\ -1^-$&$0^-,\ -\infty$\\\cline{2-3}
&$-1^+,\ \sqrt{2}$&$+\infty,\ 0^+$\\\hline
3&$-\sqrt{2},\ +1^-$&$0^-,\ -\infty$\\\cline{2-3}
&$+1^+,\ \sqrt{2}$&$+\infty,\ 0^+$\\\hline
4&$-\sqrt{2},\  \ \sqrt{2}$&$0^-,\  \  0^-$\\\hline
5&$-\sqrt{2},\ -1^-$&$0^+,\ +\infty$\\\cline{2-3}
&$-1^+,\ \sqrt{2}$&$-\infty,\ 0^-$\\\hline
6&$-\sqrt{2},\ +1^-$&$0^+,\ +\infty$\\\cline{2-3}
 &$+1^+,\ \sqrt{2}$&$-\infty,\ 0^-$\\\hline
\end{tabular}
\end{center}

We choose Branch 2 to illustrate the figure or the table. For Branch 2, $T\to 0^-$ as  $z\to-\sqrt{2}$, $T\to-\infty$ as $z\to-1^-$, $T\to+\infty$ as $z\to-1^+$ and  $T\to\sqrt{2}$ as $T\to0^+$. While in the range $-\sqrt{2}<z<-1$, $T$ is always negative, and in the range $-1<z<\sqrt{2}$, $T$ is always positive. Hence according to the signature of the temperature, we can classify the branches in more detail. The results are

\begin{center}
\begin{tabular}{|c|c|c|c|c|c|c|}
\hline
Branch&$(1,+)$&$(1,-)$&$(2,-)$&$(2,+)$&$(3,-)$&$(3,+)$\\\hline
z&$[-\sqrt{2},\sqrt{2}]$&$\phi$&$[-\sqrt{2},-1]$&$[-1,\sqrt{2}]$&$[-\sqrt{2},1]$&$[1,\sqrt{2}]$\\\hline
Branch&$(4,+)$&$(4,-)$&$(5,+)$&$(5,-)$&$(6,+)$&$(6,-)$\\\hline
z&$\phi$&$[-\sqrt{2},\sqrt{2}]$&$[-\sqrt{2},-1]$&$[-1,\sqrt{2}]$&$[-\sqrt{2},1]$&$[1,\sqrt{2}]$\\\hline
\end{tabular}
\end{center}

where the first number in the bracket denotes the branch according to the entropy function, while the second denotes the
signature of the temperature. $\phi$ means that there the value of $z$ is empty, namely, the corresponding branch does not exist.

Let us determine which branch is physically allowed, using the following two conditions:

(1) The entropy $S\ge0$;

(2) The temperature $T\ge0$.

Hence there are only three  physically allowed branches: Branch $(1,+)$, $(2,+)$, $(6,+)$. Branch (1,+) is charge conjugate invariant, while Branch (2,+) (6,+) are charge conjugate to each other. One can also determine the signature of the chemical potential in these branches by combining the  $\mu T-z$ diagram and the $T/\sqrt{\mathcal{L}}-z$ diagram. We use the symbol (Branch, Temperature, Chemical Potential)
 to characterize the branch, the temperature $T$ and the chemical potential $\mu$ in each branch.  The results are
\begin{center}
\begin{tabular}{|c|c|c|c|c|c|c|}
\hline
Branch&(1,+,-)&(1,+,+)&(2,-,-)&(2,+,+)&(3,-,-)&(3,+,+)\\\hline
z&$[-\sqrt{2},0]$&$[0,\sqrt{2}]$&$[-\sqrt{2},-1]$&$[-1,\sqrt{2}]$&$[-\sqrt{2},1]$&$[1,\sqrt{2}]$\\\hline
Branch&(4,-,-)&(4,-,+)&(5,+,-)&(5,-,+)&(6,+,-)&(6,-,+)\\\hline
z&$[-\sqrt{2},0]$&$[0,\sqrt{2}]$&$[-\sqrt{2},-1]$&$[-1,\sqrt{2}]$&$[-\sqrt{2},1]$&$[1,\sqrt{2}]$\\\hline
\end{tabular}
\end{center}

Note that when the authors plotted their phase diagram in \cite{David:2012}, they have assumed that $T>0,\mu>0$, hence, the branches they found only correspond to Branch (1,+,+),(2,+,+),(3,+,+) according to our classification. More precisely, the Branch I and II in their paper correspond to our Branch (1,+,+), Branch III is our Branch (2,+,+), Branch IV is our Branch (3,+,+).
Motivated by this, we complete the diagram by including the negative temperature
part (while we still set $\mu=1$).

\item The $\mathcal{L}-T$ diagram with $\mu=1$ is shown in Fig. \ref{L-TSpin3}.

\begin{figure}
\centering
\includegraphics[height=9cm,width=12cm]{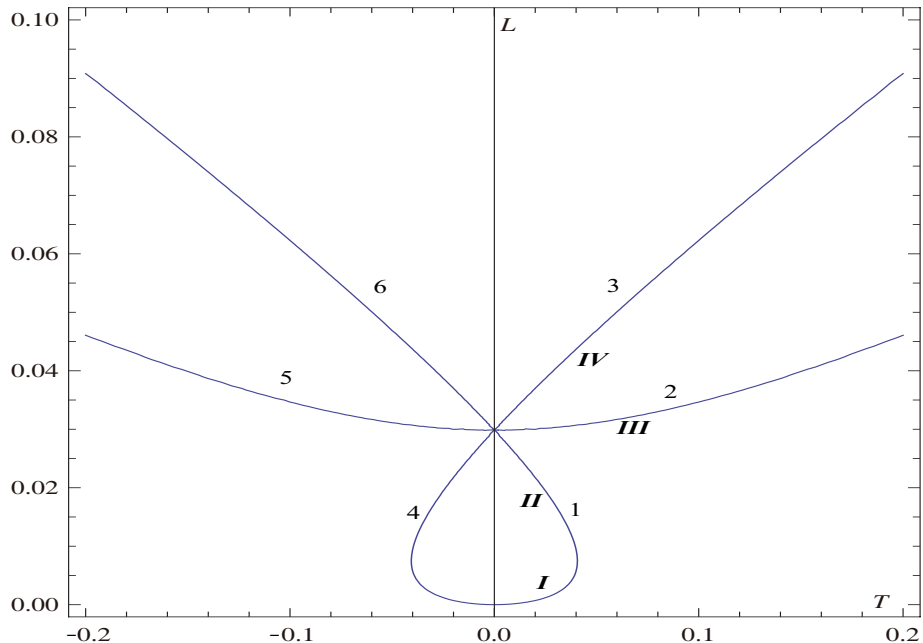}
\caption{$\mathcal{L}-T$ relation with $\mu=1$}\label{L-TSpin3}
\end{figure}
Note that when $T=0$, there are two different $\mathcal{L}$ values: $\mathcal{L}=0$ and $\mathcal{L}>0$. The one with $T=0, \mathcal{L}=0$ corresponds to the usual BTZ branch, or Branch I in \cite{David:2012}. On the contrary, the $T=0,\mathcal{L}\not=0$ is quite special. There are six curves spreading out from it, corresponding to  six branches we found from the holonomy equations. In the $T>0$ region, they are Branch 1,2,3 in an anticlockwise direction.  In the $T<0$ region, they are Branch 6,5,4 in an anticlockwise direction. In particular Branch 1 connects two points $T=0, \mathcal{L}=0$ and $T=0,\mathcal{L}>0$.  The correspondence of our classification and the classification in \cite{David:2012} is
\be
(1\leftrightarrow I,II),\  (2\leftrightarrow III),\  (3\leftrightarrow IV).
\ee
We have checked that the low temperature expansion and the high temperature expansion are exactly the
same as the ones in \cite{David:2012}. 

\item Let us turn to the phase transition. As we have shown, Branch 1 only exists in the low temperature. At the critical temperature, it has to transit to another phase. Here we only focus on the positive chemical potential $\mu>0$ as the negative $\mu$ case can be found by charge conjugation. We have found that the critical temperature is at
\be
\mu T=\frac{3}{16\pi}\sqrt{2\sqrt{3}-3}.\label{cri}
\ee
 The phase transition is from Branch (1,+,+) to Branch (2,+,+). We use the value (\ref{cri}) and the equation (\ref{mut}) in Branch (2,+,+) to find the corresponding $z$ in Branch (2,+,+). It is
\be
z=\frac{17 + 3 \sqrt{3} -
 14 \sqrt{-5 + 3 \sqrt{3}}}{\sqrt{2} (5 - 2 \sqrt{-5 + 3 \sqrt{3}})^{3/2}}\approx1.35537.
\ee
Then the entropy changes discontinuously at the critical point. The difference on the entropy is
\be
\Delta S=S_2-S_1\approx-196.777kT_c.
\ee
Here we have included the contribution from anti-holomorphic or right-moving part.
\end{enumerate}

According to the above analysis, we finally find that the full phase structure of the spin 3 black hole is
\begin{enumerate}
\item For a fixed $\mu>0$, in the range $T<T_c$, the stable part of Branch (1,+,+) or Branch I in \cite{David:2012} is thermodynamically favored. At $T=T_c$, Branch (1,+,+) transits to Branch (2,+,+). The entropy is reduced by $196.777kT_c$.
\item For fixed $\mu<0$, in the range of $T<T_c$, the stable part of Branch (1,+,-) is thermodynamically favored. At $T=T_c$, Branch (1,+,-)  transits to Branch (6,+,-). The change of the entropy is the same as $\mu>0$ case.
\end{enumerate}

In summary, we reproduced successfully the phase structure of the spin 3 black hole in \cite{David:2012}, from a different point of view. In \cite{David:2012}, the analysis rests heavily on the holonomy equations and the entropy relation. By analyzing the second holonomy equation, the authors in \cite{David:2012} did the low temperature and high temperature expansions and found the phase structure. Different from their treatments, we showed that
the entropy functions of all branches could be obtained exactly, if we do not impose the boundary condition. The entropy functions and the entropy relation (\ref{entropyrelation3}) allow us to analyze every branch in more details. This method will be more fruitful in dealing with the spin  $\tilde{4}$ black hole in next section.

\section{Phase Structure in Spin $\tilde{4}$ Black hole}

This section is to explore the phase structure of the spin $\tilde{4}$ black hole which was discovered in \cite{truncated}. We first use the method in \cite{David:2012} to explore some aspects of the holonomy equations.  Then we switch to the exact solutions to find complete picture. For simplicity, we only focus on the non-rotating black hole case.

\subsection{A First Glance of Holonomy Equations}

We start from the holonomy equations worked out in \cite{truncated}\footnote{Here we use the same convention as \cite{truncated}.}
\bea
 -20\pi^2&=&\frac{16}{5} \pi^2 (36 \alpha^2 (4 \mathcal{L}^3 + 7 \mathcal{L} \mathcal{W}) + 360 \alpha \mathcal{W} \tau +
    25 \mathcal{L} \tau^2),\\
 164\pi^4&=&\frac{64}{625} \pi^4 (1296 \alpha^4 (656 \mathcal{L}^6 + 4024 \mathcal{L}^4 \mathcal{W} + 4025 \mathcal{L}^2 \mathcal{W}^2 -
      22500 \mathcal{W}^3) \nn\\
   &&+51840 \alpha^3 \mathcal{L} (8 \mathcal{L}^4 + 82 \mathcal{L}^2 \mathcal{W} - 25 \mathcal{W}^2) \tau +
   48600 \alpha^2 (4 \mathcal{L}^4 - 9 \mathcal{L}^2 \mathcal{W} + 100 \mathcal{W}^2) \tau^2 \nn\\
   &&-36000 \alpha \mathcal{L} (2 \mathcal{L}^2 - 17 \mathcal{W}) \tau^3 + 625 (41 \mathcal{L}^2 - 36 \mathcal{W}) \tau^4)
\eea
where we have omit the subscript $4$ in $\mathcal{W}_4$ to simplify notation. Note that we can solve out $\mathcal{W}$ in terms of $\tau,\alpha,\mathcal{L}$ from the first holonomy equation and then substitute it into the second holonomy equation, making use of the non-rotating condition $\tau=\frac{i}{2\pi T}$ and $\alpha=-\mu \tau$ at the same time. We find
\bea
&&-584755200 \mu^6 \mathcal{L}^7 + 1672151040 \mu^7 \mathcal{L}^8 + 1719926784 \mu^8 \mathcal{L}^9 -
   57600 \mu^5 \mathcal{L}^6 (13405 + 17316 \mu \pi^2 T^2) \nn\\
&&   +12500 \mu^3 \mathcal{L}^4 (10724 + 21667 \mu \pi^2 T^2) -
   500 \mu^4 \mathcal{L}^5 (-210725 + 824544 \mu \pi^2 T^2) \nn\\
&&  - 78125 \pi^2 T^2 (20 - 28 \mu \pi^2 T^2 + 125 \mu^2 \pi^4 T^4) -
   15625 \mathcal{L} (-100 - 420 \mu \pi^2 T^2 + 919 \mu^2 \pi^4 T^4) \nn\\
 &&  +12500 \mu \mathcal{L}^2 (-700 - 1700 \mu \pi^2 T^2 + 1351 \mu^2 \pi^4 T^4) \nn\\
&&  + 7500 \mu^2 \mathcal{L}^3 (-2350 + 5075 \mu \pi^2 T^2 +
      23676 \mu^2 \pi^4 T^4) = 0\label{LT}
      \eea

We list some properties read from the equation (\ref{LT}) as following
\begin{enumerate}
\item The equation (\ref{LT}) is invariant under $T\to-T$ as the spin 3 gravity.
\item The equation (\ref{LT}) is asymmetric under $\mu\to-\mu$. This is a reflection of the charge conjugate asymmetry found in \cite{truncated}.
\item There are nine roots of (\ref{LT}).  We find that at $T=0$,
the equation of motion is
\be
\mathcal{L}(-25 + 35\mu \mathcal{L} + 144 \mu^2 \mathcal{L}^2)^4=0.
\ee
Thus there should be nine roots of $\mathcal{L}$, one single root $\mathcal{L}=0$ and two quadruply degenerate roots $\mathcal{L}=\frac{5}{16\mu}$ and $\mathcal{L}=-\frac{5}{9\mu}$, indicating nine branches.
However, we find that an arbitrary small perturbation around $T=0$ will render the real roots of $\mathcal{L}$ to five. The real roots depend on the signature of the chemical potential $\mu$. This will be clear later when we do low temperature expansion of $\mathcal{L}$ in terms of $T$.
\item From the equation (\ref{LT}), we can plot the $\mathcal{L}-T$ diagram. Due to the asymmetry of $\mu\to-\mu$, we need to consider positive $\mu$ and negative $\mu$ separately. For positive $\mu$, we set $\mu=1$, the $\mathcal{L}-T$ diagram is shown in Fig. \ref{L-TSpin4}.
    \begin{figure}
    \centering
\includegraphics[height=9cm,width=12cm]{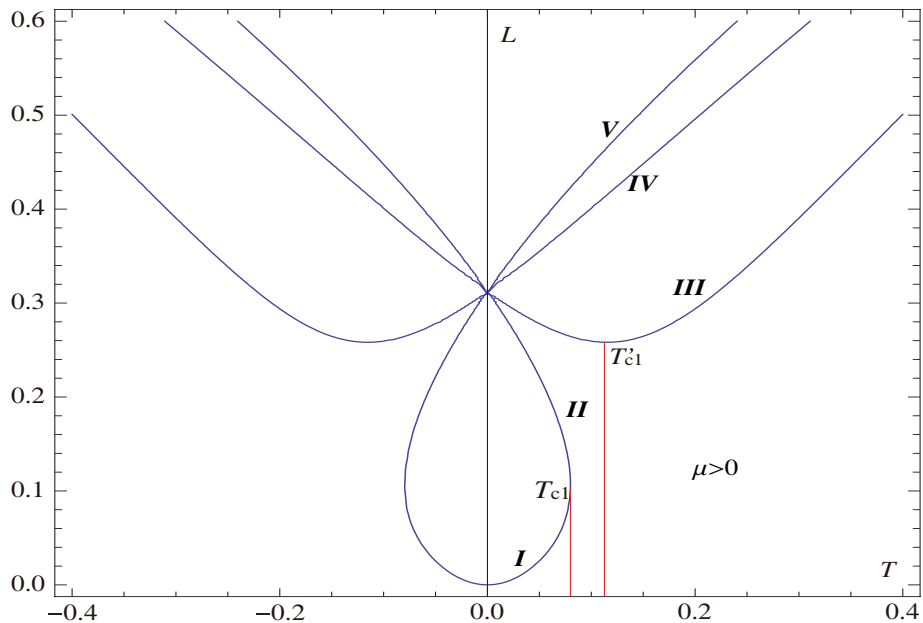}
\caption{$\mathcal{L}-T$ relation with $\mu=1$}\label{L-TSpin4}
\end{figure}
This diagram is symmetric under $T\to-T$, hence, it is sufficient to restrict oneself to the $T\ge0$ region. In this region, we can classify the curves as in \cite{David:2012}. The curve which is smoothly connected to the BTZ solution is called Branch $I$. There are four curves spread to $T>0$ region from $\mathcal{L}=\frac{5}{16\mu}$, we call them Branch $II,III,IV$ and $V$ in an anticlockwise direction. There is a critical point $T=T_{c1}$ where Branch $I,II$ merges and disappear after $T>T_{c1}$. Near $T=0$, Branch $II,III$ are unstable due to the negative specific heat. The stable branches near $T=0$ are Branch $I, IV, V$. There is another critical temperature $T'_{c1}$ above which  Branch $III$ becomes stable, too. The fact that $T_{c1}<T'_{c1}$ has important consequence in the following discussion.

For a negative $\mu$, we set $\mu=-1$, and show the $\mathcal{L}-T$ diagram in Fig. \ref{L-TSpin42}.
\begin{figure}
    \centering
\includegraphics[height=9cm,width=12cm]{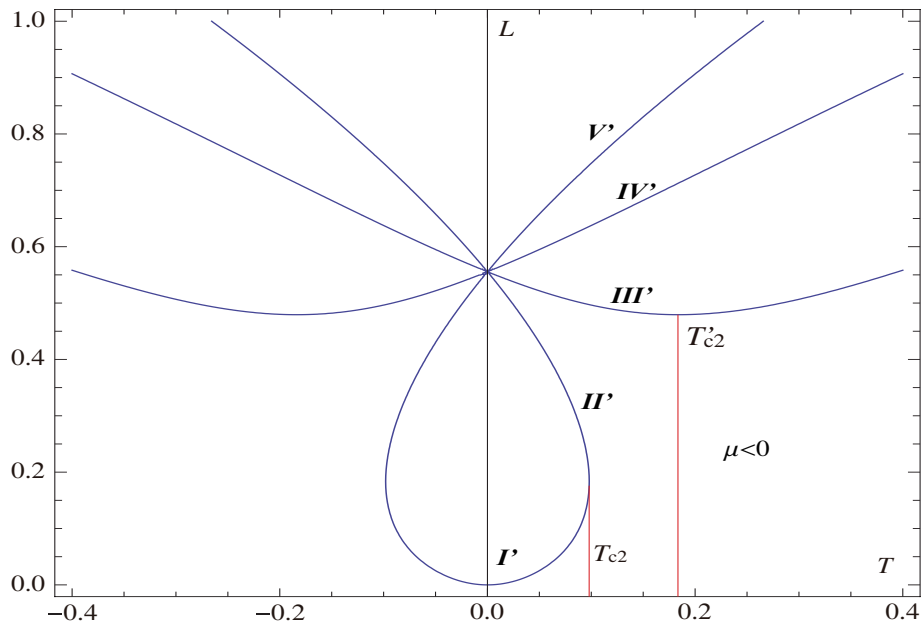}
\caption{$\mathcal{L}-T$ relation with $\mu=-1$}\label{L-TSpin42}
\end{figure}
 Again, the curve are symmetric under $T\to -T$ so we will restrict ourselves to $T>0$ region. The curve which is smoothly connected to BTZ black hole is called Branch $I'$. There are other 4 curves which are spread from $\mathcal{L}=-\frac{5}{9\mu}$ and they are called Branch $II',III', IV'$ and $V'$ in an anticlockwise direction. There is a critical point $T=T_{c2}$ where Branch $I',II'$ merges and disappear after $T>T_{c2}$. Near $T=0$, Branch $II',III'$ are unstable due to the negative specific heat. The stable Branches near $T=0$ are Branch $I', IV', V'$. Similarly there is also a critical temperature $>T'_{c2}$ above which Branch $III'$ becomes stable. From Fig. \ref{L-TSpin42}, it seems that $T_{c2}<T'_{c2}$, which will be shown explicitly soon from the exact solution. The negative $\mu$ case is much like the positive $\mu$ case, except that the character temperature is different.

 Note that whatever $\mu$ is positive or negative, there are only five branches in $T>0$. This does not contradict with the nine roots of the equation (\ref{LT}) since for arbitrary $T>0$, there are at most five real roots. The other four roots are imaginary.

\item Low temperature expansion. When $\mu>0$, we can only do low temperature expansion around $\mathcal{L}=0$ and $\mathcal{L}=\frac{5}{16\mu}$. When $\mu<0$, we can only do low temperature expansion around $\mathcal{L}=0$ and $\mathcal{L}=-\frac{5}{9\mu}$. The results are listed in the table below.

\begin{center}
\begin{tabular}{|c|c|c|}
\hline
Signature of $\mu$&Branch&Low Temperature Expansion of $\mathcal{L}$ \\\hline
/&$I,I'$&$\pi^2 T^2 (1 + \frac{1008}{25} \mu^2 \pi^4 T^4 + \frac{4032}{25} \mu^3 \pi^6 T^6
   +\cdots)$\\\hline
+&$II$&$\frac{5}{16 \m} -\frac{\pi T}{2 \sqrt{\mu}} - \frac{42\pi^2 T^2}{125} +\cdots$\\\hline
+& $III$&$\frac{5}{16 \mu} - \frac{\pi T}{4 \sqrt{\mu}} + \frac{39 \pi^2 T^2}{250} +\cdots$\\\hline
+& $IV$&$\frac{5}{16 \mu}+ \frac{\pi T}{4 \sqrt{\mu}} + \frac{39 \pi^2 T^2}{250} +\cdots$\\\hline
+& $V$&$\frac{5}{16 \m} +\frac{\pi T}{2 \sqrt{\mu}} - \frac{42\pi^2 T^2}{125} +\cdots$\\\hline
-&$II'$&$-\frac{5}{9 \mu} - \frac{\pi T}{\sqrt{-2\mu}} - \frac{54 \pi^2 T^2}{125}+\cdots$\\\hline
-&$III'$&$-\frac{5}{9 \mu}-\frac{\pi T}{3 \sqrt{-2\mu}} + \frac{14 \pi^2 T^2}{125}+\cdots$\\\hline
-&$IV'$&$-\frac{5}{9 \mu}+\frac{\pi T}{3 \sqrt{-2\mu}} + \frac{14 \pi^2 T^2}{125}+\cdots$\\\hline
-&$V'$&$-\frac{5}{9 \mu} + \frac{\pi T}{\sqrt{-2\mu}} - \frac{54 \pi^2 T^2}{125}+\cdots$\\\hline
\end{tabular}
\end{center}
Here the slash means that the signature of $\mu$ is irrelevant for the expansion around $\mathcal{L}=0$. The expansion is consistent with the previous $\mathcal{L}-T$ figure. The expansion depends on the signature of the chemical potential $\mu$. For each definite signature $\mu$, there are only five branches near $T>0$. The specific heat from the expansion also matches with the figure, indicating two unstable branches near $T>0$ for fixed $\mu$.

\item High temperature expansion. For $\mu>0$, only three branches (Branch III,IV,V) are left in the high temperature region. For $\mu<0$, only three branches (Branch III', IV', V') are left. These are consistent with the diagram plotted previously. The results are listed below

\begin{center}
\begin{tabular}{|c|c|}
\hline
Branch&High Temperature Expansion of $\mathcal{L}$ \\\hline
$V$&$\frac{(1/3 (337 - 7 \sqrt{481}))^{1/3} (5 \pi)^{2/3}}{24}(\frac{T}{\mu})^{2/3}+\cdots$\\\hline
$IV$&$\frac{(10 \pi)^{2/3}}{4*3^{2/3}}(\frac{T}{\mu})^{2/3}-\frac{35}{324\mu}+\cdots$\\\hline
$III$&$\frac{(1/3 (337 + 7 \sqrt{481}))^{1/3} (5 \pi)^{2/3}}{24}(\frac{T}{\mu})^{2/3}+\cdots$\\\hline
$III'$&$\frac{1}{24}(\frac{1}{3}(337-7\sqrt{481}))^{1/3}(5\pi)^{2/3}(\frac{T}{-\mu})^{2/3}+\cdots$\\\hline
$IV'$&$\frac{(10 \pi)^{2/3}}{4*3^{2/3}}(\frac{T}{-\mu})^{2/3}-\frac{35}{324\mu}+\cdots$\\\hline
$V'$&$\frac{1}{24}(\frac{1}{3}(337+7\sqrt{481}))^{1/3}(5\pi)^{2/3}(\frac{T}{-\mu})^{2/3}+\cdots$\\\hline
\end{tabular}
\end{center}

Actually, to determine the high temperature expansion of each branch, we need to include the $\mathcal{L}-T$ diagram in a higher temperature to keep track of the tendency of each branch. For $\mu>0$, the diagram is shown in Fig. \ref{L-TSpin43}.
\begin{figure}
    \centering
\includegraphics[height=9cm,width=12cm]{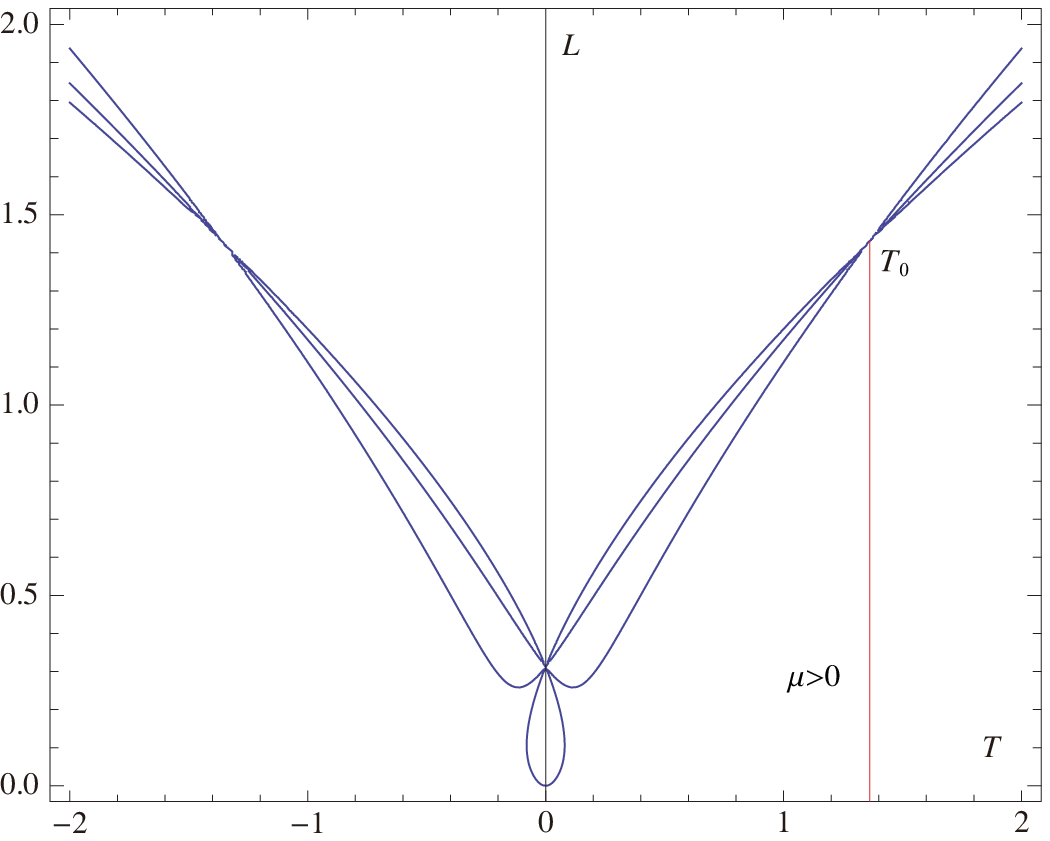}
\caption{$\mathcal{L}-T$ relation with $\mu=1$}\label{L-TSpin43}
\end{figure}

An interesting fact is that at the point $\mu T^2=\mu T_0^2$, Branch $III,IV,V$ intersect. When $T<T_0$,
\be
\mathcal{L}_{III}<\mathcal{L}_{IV}<\mathcal{L}_{V}
\ee
and when $T>T_0$, the previous relation reverses
\be
\mathcal{L}_{III}>\mathcal{L}_{IV}>\mathcal{L}_{V}.
\ee
We also notice that in the high temperature region, the relation between
the slopes $\tan\phi$ of $\mathcal{L}-T$ curve of three branches satisfy
\be
\tan\phi_{III}>\tan\phi_{IV}>\tan\phi_{V}.
\ee
This relation is consistent with the high temperature expansion of $\mathcal{L}$, as the coefficient in Branch $III$ is the largest in the expansion. The value of $T_0$ will be determined from the exact solution.  We do not find the similar behavior for $\mu<0$.

\item Note that in the low temperature region, only Branch I has the desired CFT behavior, namely, $\mathcal{L}\sim T^2$. In the high temperature region, Branch III,IV and V are left, and the high temperature behavior is $\mathcal{L}\sim T^{2/3}$, not quite the typical behavior of a CFT. In \cite{David:2012}, this puzzle was solved by RG flow to a UV CFT. In the spin $\tilde{4}$ case, this again indicates that the IR CFT should flow to another UV CFT. We will not tackle this problem in this paper.

\end{enumerate}

\subsection{Exact Solution}

Again, we find that the holonomy equations can be solved exactly as in the spin 3 gravity. We just list the results below. There are eight branches of the solutions. The entropies in each branch are respectively
\bea
S_1&=&2\pi k\sqrt{\mathcal{L}}\cos\frac{1}{4}(\arcsin{\frac{1}{25}(7-72z)}-\arcsin{\frac{7}{25}}),\\
S_2&=&2\pi k\sqrt{\mathcal{L}}\cos\frac{1}{4}(\arcsin{\frac{1}{25}(7-72z)}-\arcsin{\frac{7}{25}}+2\pi),\\
S_3&=&2\pi k\sqrt{\mathcal{L}}\cos\frac{1}{4}(\arcsin{\frac{1}{25}(7-72z)}-\arcsin{\frac{7}{25}}+4\pi),\\
S_4&=&2\pi k\sqrt{\mathcal{L}}\cos\frac{1}{4}(\arcsin{\frac{1}{25}(7-72z)}-\arcsin{\frac{7}{25}}+6\pi),\\
S_5&=&2\pi k\sqrt{\mathcal{L}}\cos\frac{1}{4}(\arcsin{\frac{1}{25}(7-72z)}+\arcsin{\frac{7}{25}}+\pi),\\
S_6&=&2\pi k\sqrt{\mathcal{L}}\cos\frac{1}{4}(\arcsin{\frac{1}{25}(7-72z)}+\arcsin{\frac{7}{25}}+3\pi),\\
S_7&=&2\pi k\sqrt{\mathcal{L}}\cos\frac{1}{4}(\arcsin{\frac{1}{25}(7-72z)}+\arcsin{\frac{7}{25}}+5\pi),\\
S_8&=&2\pi k\sqrt{\mathcal{L}}\cos\frac{1}{4}(\arcsin{\frac{1}{25}(7-72z)}+\arcsin{\frac{7}{25}}+7\pi).
\eea
 Here $z\equiv \frac{\mathcal{W}}{\mathcal{L}^2}$ to match the result in \cite{truncated}. It is dimensionless and is in the range $[-\frac{1}{4},\frac{4}{9}]$. Note that to obtain these results we do not use the non-rotating condition and we include only the holomorphic part contribution.
The eight branches are actually the eight curves that spread from $\mathcal{L}=\frac{5}{16\mu}$ or $\mathcal{L}=-\frac{5}{9\mu}$.
From the exact solutions, we can study the phase structure more precisely than before.

\begin{enumerate}
\item Here we present all the relations that are necessary in the discussion below. $\tau$ and $\alpha$ are
\bea
\tau&=&\frac{i}{2\sqrt{\mathcal{L}}}(\cos\th(z)-\frac{6z\sin\th(z)}{\sqrt{(1+4z)(4-9z)}})\equiv\frac{i}{2\sqrt{\mathcal{L}}}g(z),\\
\alpha&=&\frac{5i}{12\mathcal{L}^{3/2}}\frac{\sin\th(z)}{\sqrt{(1+4z)(4-9z)}}\equiv\frac{5i}{12\mathcal{L}^{3/2}}h(z).
\eea
In the non-rotating case, we have
\bea
\mathcal{L}\mu&=&-\frac{5}{6}\frac{h(z)}{g(z)},\\
\mu T^2&=&-\frac{5}{6\pi^2}\frac{h(z)}{g(z)^3},\la{muT2}\\
S&=&2\pi^2 k T g(z)\cos\th(z),\\
\mu S^2&=&-\frac{10\pi^2k^2}{3}\frac{h(z)}{g(z)}\cos^2\th(z).
\eea
One can check that the relation (\ref{entropy}) as\footnote{Note that the coefficients $\frac{k}{2\pi}$ and $\frac{9k}{5\pi}$ originate from the normalization convention in \cite{truncated}.}
\be
S=-4\pi^2 i(2\tau\times\frac{k}{2\pi}\mathcal{L}+4\alpha\times\frac{9k}{5\pi}\mathcal{W}).
\ee
\item Again, we plot the diagram $S/S_{0}-z$ to show the discrepancy of each branch from the $BTZ$ black hole. The figure is shown in Fig. \ref{szSpin4}.
\begin{figure}
\centering
\includegraphics[height=9cm,width=12cm]{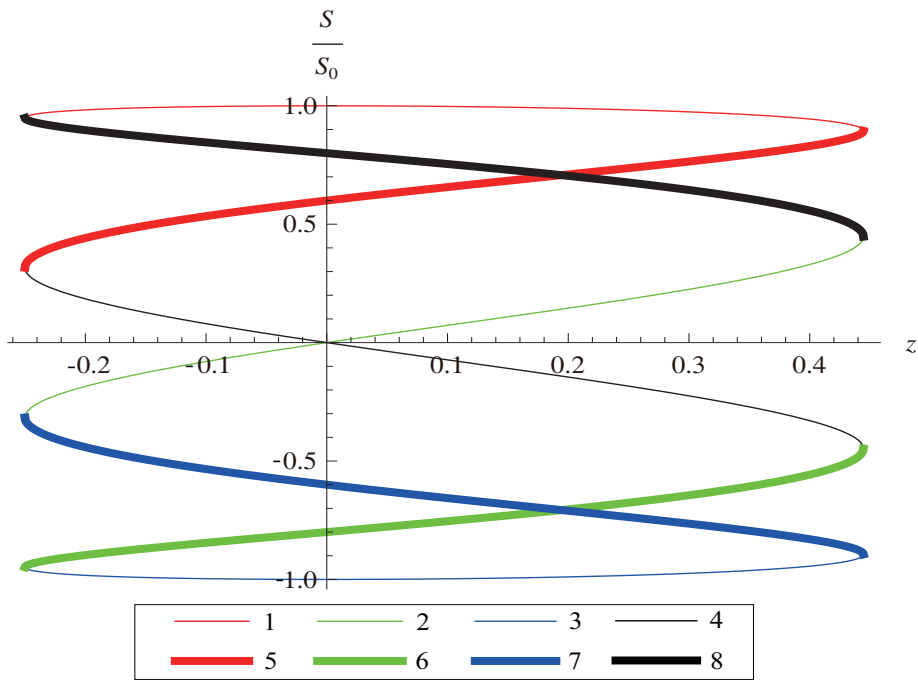}
\caption{$S/S_{0}-z$ relation}\label{szSpin4}
\end{figure}
Since there are eight branches now, we use the color and thickness of the curve to distinguish them. From Branch 1 to Branch 8, they are depicted by the curves in
\bea
\mbox{(Red,Thin),(Green,Thin),(Blue,Thin),(Black,Thin),}\nn\\
\mbox{(Red,Thick),(Green,Thick),(Blue,Thick),(Black,Thick).}\nn
\eea
 Unlike the spin 3 case, Branch 2 and 4 may have positive or negative entropy, depending on the range of $z$. We distinguish them by Branch $2_1,2_2,4_1$ and $4_2$, in which the subscript $1$ means that $z\in[-\frac{1}{4},0]$ and the subscript $2$ means that $z\in[0,\frac{4}{9}]$. Hence the signature in each branch is
\be
(1,+),(2_1,-),(2_2,+),(3,-),(4_1,+),(4_2,-),(5,+),(6,-),(7,-),(8,+).
\ee
The requirement that the entropy $S\ge0$  tell us that only Branch $1,2_2,4_1,5$ and $8$ are physically allowed.

\item We can plot the $T/\sqrt{\mathcal{L}}-z$ diagram. The figure is shown in Fig. \ref{TzSpin4}.
\begin{figure}
\centering
\includegraphics[height=9cm,width=12cm]{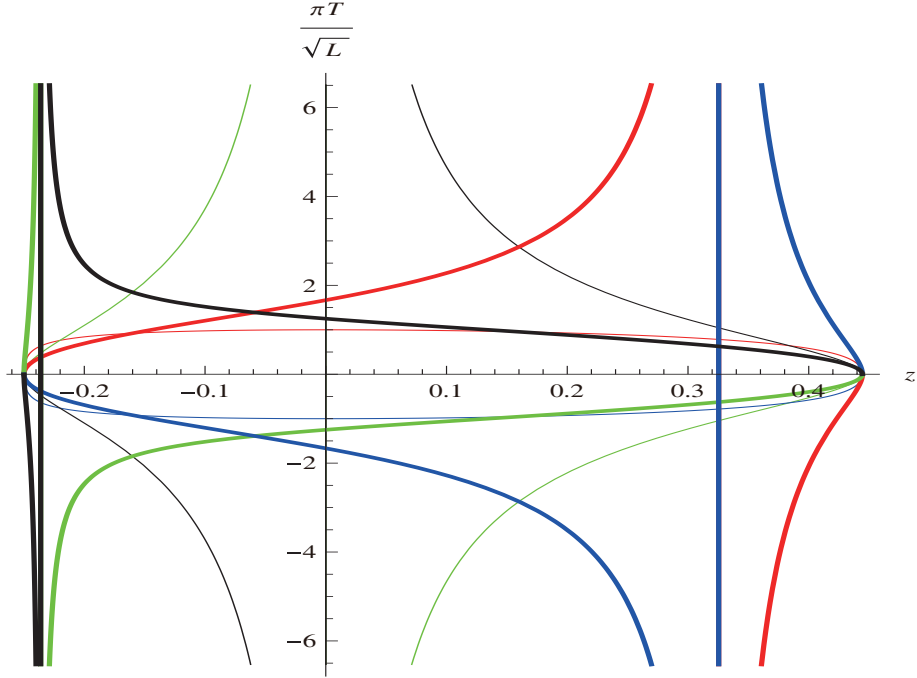}
\caption{$T/\sqrt{\mathcal{L}}-z$ relation}\label{TzSpin4}
\end{figure}
The figure is more complicated than the spin 3 case, but the qualitative behavior is the same. We can  summarize the signatures of the temperatures in each branch in a table

\begin{center}
\begin{tabular}{|c|c|c|}
\hline
Branch&z&T\\\hline
1&$-1/4,\  \ 4/9$&$0^+,\  \  0^+$\\\hline
$2_1$&$-1/4,\ 0^-$&$0^+,\ +\infty$\\\hline
$2_2$&$0^+,\ 4/9$&$-\infty,\ 0^-$\\\hline
3&$-1/4,\  \ 4/9$&$0^-,\  \  0^-$\\\hline
$4_1$&$-1/4,\ 0^-$&$0^-,\ -\infty$\\\hline
$4_2$&$0^+,\ 4/9$&$+\infty,\ 0^+$\\\hline
5&$-1/4,\ z_1^-$&$0^+,\ +\infty$\\\cline{2-3}
&$z_1^+,\ 4/9$&$-\infty,\ 0^-$\\\hline
6&$-1/4,\ z_2^-$&$0^+,\ +\infty$\\\cline{2-3}
&$z_2^+,\ 4/9$&$-\infty,\ 0^-$\\\hline
7&$-1/4,\ z_1^-$&$0^-,\ -\infty$\\\cline{2-3}
&$z_1^+,\ 4/9$&$+\infty,\ 0^+$\\\hline
8&$-1/4,\ z_2^-$&$0^-,\ -\infty$\\\cline{2-3}
&$z_2^+,\ 4/9$&$+\infty,\ 0^+$\\\hline
\end{tabular}
\end{center}

The temperature  is always positive in Branch 1, and always negative in Branch 3. In Branch 2 and 4, it flips a sign at $z=z_0=0$. In Branch 5 and 7, it flips a sign at $z=z_1=0.325526$. In Branch 6 and 8, it flips a sign at $z=z_2=-0.235926$. Here $z_0,z_1,z_2$ are determined by solving the equation
\be
g(z)\equiv\cos\th(z)-\frac{6z\sin\th(z)}{\sqrt{(1+4z)(4-9z)}}=0.
\ee
The solution is
\be
z=z_0=0
\ee
for Branch 2 and 4,
\be
z=z_1=\frac{4(\sqrt{481}-7)}{337-7\sqrt{481}}\approx0.325526
\ee
for Branch 5 and 7, and
\be
z=z_2=\frac{-4(\sqrt{481}+7)}{337+7\sqrt{481}}\approx-0.235926
\ee
for Branch 6 and 8.
There is no real solution for Branch 1 and 3, consistent with the graph in Fig. \ref{TzSpin4}.
Hence, according to the signatures of the temperatures, we distinguish the previous branches  more carefully as

\begin{center}
\begin{tabular}{|c|c|c|c|c|c|c|c|}
\hline
Branch&$(1,+)$&$(2_1,+)$&$(2_2,-)$&$(3,-)$&$(4_1,-)$&$(4_2,+)$&$(5,+)$\\\hline
z&$[-\frac{1}{4},\frac{4}{9}]$&$[-\frac{1}{4},0]$&$[0,\frac{4}{9}]$&$[-\frac{1}{4},\frac{4}{9}]$&$[-\frac{1}{4},0]$&$[0,\frac{4}{9}]$&$[-\frac{1}{4},z_1]$\\\hline
Branch&$(5,-)$&$(6,+)$&$(6,-)$&$(7,-)$&$(7,+)$&$(8,-)$&$(8,+)$\\\hline
z&$[z_1,\frac{4}{9}]$&$[-\frac{1}{4},z_{2}]$&$[z_{2},\frac{4}{9}]$&$[-\frac{1}{4},z_1]$&$[z_1,\frac{4}{9}]$&$[-\frac{1}{4},z_2]$&$[z_2,\frac{4}{9}]$\\\hline
\end{tabular}
\end{center}

If we require $T\ge0$, then only seven of the fourteen branches are allowed. Including the constraints from the entropy $S\ge0$, only Branch (1,+),(5,+),(8,+) are physically allowed.

\item It is also important to plot the $\mu T^2-z$ diagram from which we can read out the critical point that Branch 1 disappear. From the equation (\ref{muT2}), we find that
\be
\mu T^2=-\frac{5}{6\pi^2}\frac{\sin\th(z)/\sqrt{(1+4z)(4-9z)}}{(\cos\th(z)-6z\sin\th(z)/\sqrt{(1+4z)(4-9z)})^3}.
\ee
We notice that the angel $\th(z)$ of Branch 1 and 3 are related by $\th_1(z)\to\th_3(z)=\th_1(z)+\pi$, so the function $\mu T^2$ is the same in both branches. The same property exists between Branch 2 and 4, 5 and 7, 6 and 8. So we only need to plot four different curves. The four curves are plotted by four different color. The correspondence between the curves and the color is
\be
(1,3 \leftrightarrow \mbox{Red}),\ (2,4\leftrightarrow \mbox{Green}),\ (5,7\leftrightarrow \mbox{Blue}),\ (6,8\leftrightarrow \mbox{Black}).
\ee
Then the relations between $\mu T^2$ and $z$  are shown in Fig. \ref{T2zSpin4}.
\begin{figure}
\centering
\includegraphics[height=9cm,width=12cm]{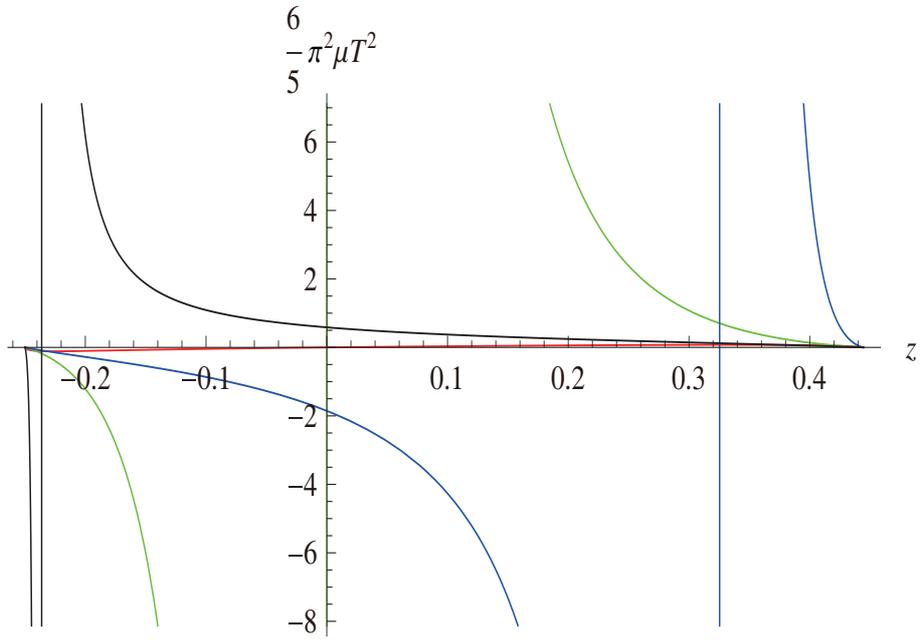}
\caption{$\mu T^2-z$ relations}\label{T2zSpin4}
\end{figure}

Note that there are still three important values of $z$. They are the same as before $z=0,z_1,z_2$. Since the red curve for Branch 1 and 3 can not be seen clearly in this figure, we single it out in Fig. \ref{T2zSpin42}.
\begin{figure}
\centering
\includegraphics[height=9cm,width=12cm]{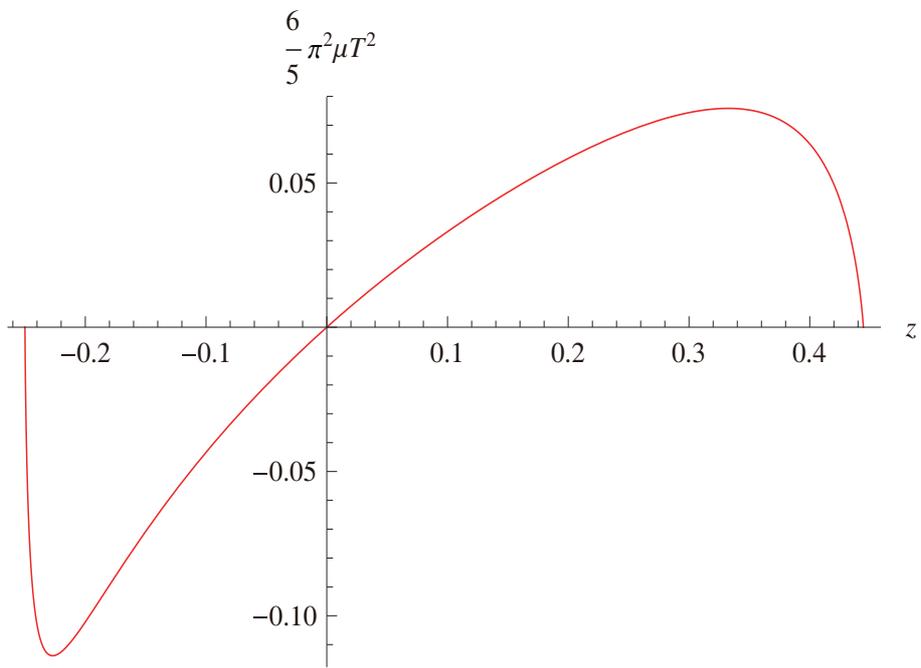}
\caption{$\mu T^2-z$ relations for Branch 1 and 3}\label{T2zSpin42}
\end{figure}

Similar to the spin 3 gravity, we can further distinguish the branches according to the signature of the chemical potential. The result is shown in the following table.

\begin{center}
\begin{tabular}{|c|c|c|c|c|c|c|c|c|}
\hline
Branch&(1,+,-)&(1,+,+)&$(2_1,+,-)$&$(2_2,-,+)$&(3,-,-)&(3,-,+)&$(4_1,-,-)$&$(4_2,+,+)$\\\hline
z&$[-\frac{1}{4},0]$&$[0,\frac{4}{9}]$&$[-\frac{1}{4},0]$&$[0,\frac{4}{9}]$&$[-\frac{1}{4},0]$&$[0,\frac{4}{9}]$&$[-\frac{1}{4},0]$&$[0,\frac{4}{9}]$\\\hline
Branch&(5,+,-)&(5,-,+)&(6,+,-)&(6,-,+)&(7,-,-)&(7,+,+)&(8,-,-)&(8,+,+)\\\hline
z&$[-\frac{1}{4},z_1]$&$[z_1,\frac{4}{9}]$&$[-\frac{1}{4},z_2]$&$[z_2,\frac{4}{9}]$&$[-\frac{1}{4},z_1]$&$[z_1,\frac{4}{9}]$&$[-\frac{1}{4},z_2]$&$[z_2,\frac{4}{9}]$\\\hline
\end{tabular}
\end{center}

Like the spin 3 gravity, there are some branches that can have vanishing spin 4 charge but nonvanishing chemical potential. The values of $\mu T^2$ at the point $z=0$ in each branch are respectively
\be
(1,0),\ (2,\infty),\ (3,0),\ (4,\infty),\ (5,-\frac{125}{81\pi^2}),\ (6,\frac{125}{256\pi^2}),\ (7,-\frac{125}{81\pi^2}),\ (8,\frac{125}{256\pi^2})
\ee

\item In the previous subsection, we found five branches in $T>0$ region for each fixed $\mu$. It is important to see the connection of the branches found here and the previous ones. This can be done by doing low and high temperature expansions of $\mathcal{L}$. For the convenience of the following discussions, we also include the expansion of the entropy. The results are listed in the table below


\begin{center}
\begin{tabular}{|c|c|c|}
\hline
Branch&z&Low T Expansion of $\mathcal{L}$ \\\hline
1&0&$\pi^2 T^2 + \frac{1008}{25} \mu^2 \pi^6 T^6 + \frac{4032}{25} \mu^3 \pi^8 T^8
   +\cdots$\\\hline
3&0&$\pi^2 T^2 + \frac{1008}{25} \mu^2 \pi^6 T^6 + \frac{4032}{25} \mu^3 \pi^8 T^8
   +\cdots$\\\hline
1&4/9&$\frac{5}{16\mu}-\frac{\pi T}{2\sqrt{\mu}}-\frac{42\pi^2 T^2}{125}+\cdots$\\\hline
2&4/9&$\frac{5}{16 \mu} -\frac{\pi T}{4 \sqrt{\mu}} +\frac{39\pi^2 T^2}{250} +\cdots$\\\hline
3 &4/9&$\frac{5}{16 \mu} + \frac{\pi T}{2 \sqrt{\mu}} - \frac{42 \pi^2 T^2}{125} +\cdots$\\\hline
4 &4/9&$\frac{5}{16 \mu}+ \frac{\pi T}{4 \sqrt{\mu}} + \frac{39 \pi^2 T^2}{250} +\cdots$\\\hline
5 &4/9&$\frac{5}{16 \mu} -\frac{\pi T}{2 \sqrt{\mu}} - \frac{42\pi^2 T^2}{125} +\cdots$\\\hline
6&4/9&$\frac{5}{16 \mu} + \frac{\pi T}{4\sqrt{\mu}} + \frac{39 \pi^2 T^2}{250}+\cdots$\\\hline
7&4/9&$\frac{5}{16 \mu}+\frac{\pi T}{2 \sqrt{\mu}}- \frac{42 \pi^2 T^2}{125}+\cdots$\\\hline
8&4/9&$\frac{5}{16 \mu}-\frac{\pi T}{4 \sqrt{\mu}} + \frac{39 \pi^2 T^2}{250}+\cdots$\\\hline
1&-1/4&$-\frac{5}{9 \mu} -\frac{\pi T}{\sqrt{-2\mu}} - \frac{54 \pi^2 T^2}{125}+\cdots$\\\hline
2&-1/4&$-\frac{5}{9 \mu} +\frac{\pi T}{3\sqrt{-2\mu}} +\frac{14 \pi^2 T^2}{125}+\cdots$\\\hline
3&-1/4&$-\frac{5}{9 \mu}+\frac{\pi T}{\sqrt{-2\mu}} - \frac{54 \pi^2 T^2}{125}+\cdots$\\\hline
4&-1/4&$-\frac{5}{9 \mu} -\frac{\pi T}{3\sqrt{-2\mu}} + \frac{14 \pi^2 T^2}{125}+\cdots$\\\hline
5&-1/4&$-\frac{5}{9 \mu} -\frac{\pi T}{3\sqrt{-2\mu}} + \frac{14 \pi^2 T^2}{125}+\cdots$\\\hline
6&-1/4&$-\frac{5}{9 \mu} +\frac{\pi T}{\sqrt{-2\mu}} - \frac{54 \pi^2 T^2}{125}+\cdots$\\\hline
7&-1/4&$-\frac{5}{9 \mu} +\frac{\pi T}{3\sqrt{-2\mu}} + \frac{14 \pi^2 T^2}{125}+\cdots$\\\hline
8&-1/4&$-\frac{5}{9 \mu} -\frac{\pi T}{\sqrt{-2\mu}} - \frac{54 \pi^2 T^2}{125}+\cdots$\\\hline
Branch&z&High T Expansion of $\mathcal{L}$\\\hline
4&$z_0$&$(\frac{5\pi}{12})^{2/3}(\frac{T}{\mu})^{2/3}-\frac{35}{324\mu}+\cdots$\\\hline
7&$z_1$&$\frac{1}{24}(\frac{1}{3}(337-7\sqrt{481}))^{1/3}(5\pi)^{2/3}(\frac{T}{\mu})^{2/3}+\cdots$\\\hline
8&$z_2$&$\frac{1}{24}(\frac{1}{3}(337+7\sqrt{481}))^{1/3}(5\pi)^{2/3}(\frac{T}{\mu})^{2/3}+\cdots$\\\hline
2&$z_0$&$(\frac{5\pi}{12})^{2/3}(\frac{T}{-\mu})^{2/3}-\frac{35}{324\mu}+\cdots$\\\hline
5&$z_1$&$\frac{1}{24}(\frac{1}{3}(337-7\sqrt{481}))^{1/3}(5\pi)^{2/3}(\frac{T}{-\mu})^{2/3}+\cdots$\\\hline
6&$z_2$&$\frac{1}{24}(\frac{1}{3}(337+7\sqrt{481}))^{1/3}(5\pi)^{2/3}(\frac{T}{-\mu})^{2/3}+\cdots$\\\hline
\end{tabular}
\end{center}

\begin{center}
\begin{tabular}{|c|c|c|}
\hline
Branch&z&Low T Expansion of $S/(2\pi k)$ \\\hline
1&0&$\pi T+\frac{432}{25}\mu^2 \pi^5T^5+\frac{8064}{125}\mu^3\pi^7T^7+\cdots$\\\hline
3&0&$\pi T+\frac{432}{25}\mu^2 \pi^5T^5+\frac{8064}{125}\mu^3\pi^7T^7+\cdots$\\\hline
1&4/9&$\frac{1}{2\sqrt{\mu}}-\frac{42\pi T}{125}-\frac{5868\sqrt{\mu}\pi^2 T^2}{15625}+\cdots$\\\hline
2&4/9&$\frac{1}{4\sqrt{\mu}}+\frac{39\pi T}{250}+\frac{4653\sqrt{\mu}\pi^2 T^2}{31250}+\cdots$\\\hline
3 &4/9&$-\frac{1}{2\sqrt{\mu}}-\frac{42\pi T}{125}+\frac{5868\sqrt{\mu}\pi^2 T^2}{15625}+\cdots$\\\hline
4 &4/9&$-\frac{1}{4\sqrt{\mu}}+\frac{39\pi T}{250}-\frac{4653\sqrt{\mu}\pi^2 T^2}{31250}+\cdots$\\\hline
5 &4/9&$\frac{1}{2\sqrt{\mu}}-\frac{42\pi T}{125}-\frac{5868\sqrt{\mu}\pi^2 T^2}{15625}+\cdots$\\\hline
6&4/9&$-\frac{1}{4\sqrt{\mu}}+\frac{39\pi T}{250}-\frac{4653\sqrt{\mu}\pi^2 T^2}{31250}+\cdots$\\\hline
7&4/9&$-\frac{1}{2\sqrt{\mu}}-\frac{42\pi T}{125}+\frac{5868\sqrt{\mu}\pi^2 T^2}{15625}+\cdots$\\\hline
8&4/9&$\frac{1}{4\sqrt{\mu}}+\frac{39\pi T}{250}+\frac{4653\sqrt{\mu}\pi^2 T^2}{31250}+\cdots$\\\hline
1&-1/4&$\frac{1}{\sqrt{-2\mu}}-\frac{54\pi T}{125}-\frac{17739\sqrt{-\mu}\pi^2T^2}{31250\sqrt{2}}+\cdots$\\\hline
2&-1/4&$-\frac{1}{3\sqrt{-2\mu}}+\frac{14\pi T}{125}-\frac{3033\sqrt{-\mu}\pi^2 T^2}{31250\sqrt{2}}+\cdots$\\\hline
3&-1/4&$-\frac{1}{\sqrt{-2\mu}}-\frac{54\pi T}{125}+\frac{17739\sqrt{-\mu}\pi^2T^2}{31250\sqrt{2}}+\cdots$\\\hline
4&-1/4&$\frac{1}{3\sqrt{-2\mu}}+\frac{14\pi T}{125}+\frac{3033\sqrt{-\mu}\pi^2 T^2}{31250\sqrt{2}}+\cdots$\\\hline
5&-1/4&$\frac{1}{3\sqrt{-2\mu}}+\frac{14\pi T}{125}+\frac{3033\sqrt{-\mu}\pi^2 T^2}{31250\sqrt{2}}+\cdots$\\\hline
6&-1/4&$-\frac{1}{\sqrt{-2\mu}}-\frac{54\pi T}{125}+\frac{17739\sqrt{-\mu}\pi^2T^2}{31250\sqrt{2}}+\cdots$\\\hline
7&-1/4&$-\frac{1}{3\sqrt{-2\mu}}+\frac{14\pi T}{125}-\frac{3033\sqrt{-\mu}\pi^2 T^2}{31250\sqrt{2}}+\cdots$\\\hline
8&-1/4&$\frac{1}{\sqrt{-2\mu}}-\frac{54\pi T}{125}-\frac{17739\sqrt{-\mu}\pi^2T^2}{31250\sqrt{2}}+\cdots$\\\hline
Branch&z&High T Expansion of $S/(2\pi k)$\\\hline
4&$z_0$&$-\frac{(\frac{5}{3})^{2/3}}{6(2\pi)^{1/3}}(\frac{1}{T})^{1/3}(\frac{1}{\mu})^{2/3}+\cdots$\\\hline
7&$z_1$&$-\frac{((-4879+481\sqrt{481})\pi)^{1/3}}{10*6^{2/3}}(\frac{T}{\mu})^{1/3}+\cdots$\\\hline
8&$z_2$&$\frac{((4879+481\sqrt{481})\pi)^{1/3}}{10*6^{2/3}}(\frac{T}{\mu})^{1/3}+\cdots$\\\hline
2&$z_0$&$-\frac{(\frac{5}{3})^{2/3}}{6(2\pi)^{1/3}}(\frac{1}{T})^{1/3}(\frac{1}{-\mu})^{2/3}+\cdots$\\\hline
5&$z_1$&$\frac{((-4879+481\sqrt{481})\pi)^{1/3}}{10*6^{2/3}}(\frac{T}{-\mu})^{1/3}+\cdots$\\\hline
6&$z_2$&$-\frac{((4879+481\sqrt{481})\pi)^{1/3}}{10*6^{2/3}}(\frac{T}{-\mu})^{1/3}+\cdots$\\\hline
\end{tabular}
\end{center}

Some illustrations on the results are in order:
\begin{enumerate}
\item There are three different $z =0,\frac{4}{9},-\frac{1}{4}$ where the temperature tends to zero. When $z\to0$, the temperatures in Branch 1 and 3 go to zero. When $z=\frac{4}{9}$ and $z=-\frac{1}{4}$, all the branches go to the low temperature region. Our low temperature expansion suggests that the eight curves around $\mathcal{L}=\frac{5}{16\mu}$ are Branch
\be
1,8,4,7,5,2,6,3
\ee
in an anticlockwise direction. And the correspondence of the previous classification and the classification here is
\be
(1\leftrightarrow I, II),\ (8\leftrightarrow III),\ (4\leftrightarrow IV),\ (7\leftrightarrow V)
\ee
Note that the classification in the previous subsection does not consider the $T<0$ region hence they correspond to only four of the eight branches we find here.
Similarly, the expansion around $z=-\frac{1}{4}$ tells us that the eight curves around $\mathcal{L}=-\frac{5}{9\mu}$ are Branch
\be
1,5,2,6,8,4,7,3
\ee
in an anticlockwise direction.
And the correspondence of the previous classification and the classification here is
\be
(1\leftrightarrow I',II'),\ (5\leftrightarrow III'),\ (2\leftrightarrow IV'),\ (6\leftrightarrow V')
\ee
\item There are also three different $z= z_0,z_1,z_2$ where the temperature can be infinity. In the high temperature expansion, we only include the case that $T>0$. When $z\to z_0(z_1,z_2)$, the temperatures in Branch 4(7,8) and 2(5,6) tend to $\infty$. Note that this is consistent with all the results in the last subsection.
\item In the previous discussion, we mentioned that in $\mathcal{L}-T$ diagram, there is a tri-point where Branch $III,IV,V$ intersect. Note that Branch $III,IV,V$ correspond to Branch 8,4,7, hence the point can be determined by the equations
\be
\mathcal{L}_8=\mathcal{L}_7=\mathcal{L}_4\equiv\mathcal{L}_0,\ \ T_8=T_7=T_4\equiv T_0,
\ee
where the subscript denotes Branch. These equations can be solved consistently by the solution
\be
z_8=-0.218535,\ z_7=0.380182,\ z_4=0.13002
\ee
with
\be
\mu T_0^2=\frac{6250}{343 \pi^2},\ \mathcal{L}_0=\frac{10}{7 \mu}=\frac{49 \pi^2 T_0^2}{625}.
\ee
Note that $z_8,z_7,z_4$ are the solutions of the equation
\be
\cos\th(z)=\frac{7-72z}{25}
\ee
in Branch 8,7,4 correspondingly.
\end{enumerate}
\end{enumerate}

\subsection{Phase Structure}

Here we consider two cases which are physically allowed. The first case is
\be
S\ge0,\hs{3ex}T\ge0,\hs{3ex} \mu>0.
\ee
There are only Branch (1,+,+) and (8,+,+) satisfying the conditions. Branch (1,+,+) is smoothly connected to the $BTZ$ black hole, and we can see clearly that at low temperature, $T\to0$, the entropy in Branch (1,+,+) is smaller\footnote{Here we mean the Branch (1,+,+) near $z=0$. There are actually two points($z=0$ or $z=\frac{4}{9}$) where $T=0$ in Branch (1,+,+). The entropy at $z=\frac{4}{9}$ is really larger than the one in Branch (8,+,+), but it is unstable near $z=\frac{4}{9}$.  When we consider negative $\mu$ case, we meet the same phenomenon.}  than the one in Branch (8,+,+).

\begin{figure}
\centering
\includegraphics[height=9cm]{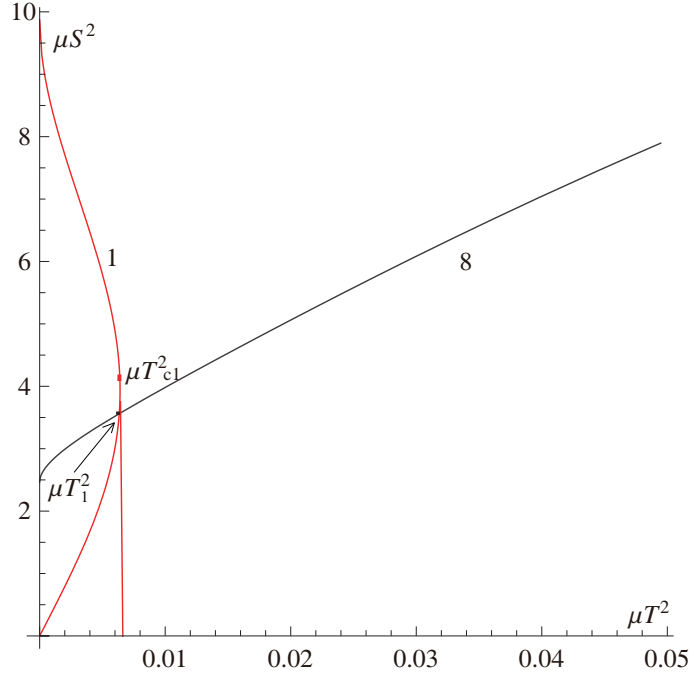}
\caption{$S-T$ relation for positive chemical potential}\label{STSpin4}
\end{figure}

However we find that the specific heat of Branch (8,+,+) is negative, indicating that Branch (8,+,+) is unstable in this temperature region. Hence, at low temperature, Branch (1,+,+) is thermodynamically favored. However, from Fig. \ref{T2zSpin42} we see that it cannot exist at arbitrary high temperature for fixed $\mu$, while Branch (8,+,+) can exist for arbitrary high temperature. Hence, there exists a critical point $T=T_{c1}$ where Branch (1,+,+) changes to Branch (8,+,+). The critical value can be determined by finding the extreme value of $\mu T^2-z$ curve. The critical value is at
\be
z=z_{c1}=0.332433,\ \mu T_{c1}^2=0.00640411.
\ee
We notice that $z_{c1}\gg z_2$. Hence when $\mu>0$, at low temperature, the system  keeps in Branch (1,+,+) until $T\to T_{c1}$, then the system undergoes a phase transition to Branch (8,+,+). We can calculate the entropies in Branch (1,+,+) and (8,+,+) at the critical temperature. They are
\be
S_1(T=T_{c1})=24.7796 k T_{c1},\ \ S_8(T=T_{c1})=23.6077 k T_{c1}.
\ee
Hence, the change of the entropy at this point is
\be
\Delta S=S_8-S_1=-1.17202 k T_{c1}.
\ee
The Fig. \ref{STSpin4}  is to show the $S-T$ relation in Branch (1,+,+) and (8,+,+).

 We use $T_1$ to denote the point where the entropies in Branch (1,+,+) and (8,+,+) equal. We can easily determine the value of $T_1$ to be
\be
\mu T_1^2=0.00633862
\ee
and the corresponding entropy
\be
S_1(T=T_1)=S_8(T=T_1)=23.7034 k T_1.
\ee

However, the picture is even subtler.  From the $\mathcal{L}-T$ diagram, we find that when Branch (1,+,+) disappears, the slope of $\mathcal{L}$ in Branch (8,+,+) is still  negative, indicating a negative specific heat. Hence at this point, Branch (8,+,+) is actually unstable. This can be checked by the low temperature expansion\footnote{Since $z_{c1}$ is far away from the high temperature point $z_2$, the low temperature expansion is robust here.} of $\mathcal{L}$ in Branch (8,+,+).  There is a turning point $T=T_{c1}'>T_{c1}$, where the specific heat of Branch (8,+,+) becomes positive. This point can be determined by the extreme point of the function
 \be
 \mathcal{L}\mu=-\frac{5}{6}\frac{\sin\th(z)/\sqrt{(1+4z)(4-9z)}}{\cos\th(z)-6z\sin\th(z)/\sqrt{(1+4z)(4-9z)}}\label{Lmu}
 \ee
 of Branch 8. The result is that the extreme point is at
 \be
 z_{c1}'=0.2874.
 \ee
 The corresponding $\mathcal{L}$ and $T$ are respectively
 \be
 \mathcal{L}=\frac{0.2582}{\mu}=19.40 T_{c1}'^2,\ \ \mu T_{c1}'^2=0.01331.
 \ee
 After $T>T_{c1}'$, it becomes stable and remains in Branch (8,+,+) to arbitrary high temperature. Note that $T_{c1}' > T_{c1}$.

\begin{figure}
\centering
\includegraphics[width=9cm]{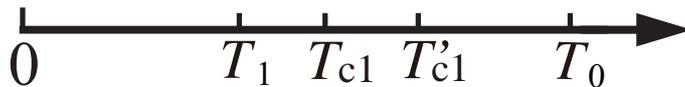}
\caption{Phase structure with positive chemical potential}\label{temp}
\end{figure}

  Therefore the full picture can be shown by Fig. \ref{temp}. When $0<T<T_1$, the system is in Branch (1,+,+) due to the fact that Branch (8,+,+) is unstable in this temperature range even though its entropy is larger. At $T=T_1$, the entropies of these two branches agree.  When $T_1<T<T_{c1}$, the system is still in Branch (1,+,+). When $T=T_{c1}^+$, Branch (1,+,+) disappears and the system is forced to Branch (8,+,+) even though it is unstable. It is unstable until $T\to T_{c1}'$, after which the specific heat becomes positive hence the system is stable. There is a higher temperature $T_0$ where Branch 4,7,8 intersect. The system remains at Branch (8,+,+) to arbitrary high temperature.

\begin{figure}
\centering
\includegraphics[height=9cm]{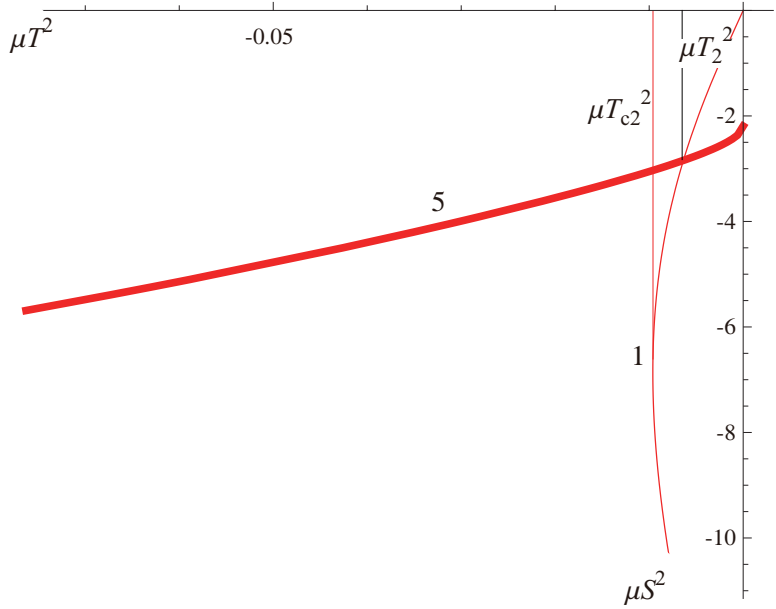}
\caption{$S-T$ relation for negative chemical potential}\label{STSpin4neg}
\end{figure}

The second case is when the chemical potential $\mu<0$
\be
S\ge0,\hs{3ex}T\ge0,\hs{3ex}\mu<0.
\ee
There are only Branch (1,+,-) and (5,+,-) satisfying the conditions. Branch (1,+,-) is smoothly connected to the $BTZ$ black hole. At low temperature, from the low temperature expansion, we find that the entropy in Branch (1,+,-) is smaller than Branch (5,+,-). However, the specific heat in Branch (5,+,-) in this region is negative so that Branch (1,+,-) is still thermodynamically favored. However, we also see from the $\mu T^2-z$ diagram that this branch cannot exist at arbitrary high temperature. Since Branch (5,+,-) is always exist, we find that there should be a critical point that Branch (1,+,-) changes to Branch (5,+,-). It is at
\be
z=z_{c2}=-0.227006,\ \mu T_{c2}^2=-0.00961277.
\ee
We notice that $z_{c2}\ll z_1$. Hence, when $\mu<0$, as the temperature grows, the system keeps in Branch (1,+,-) until $T\to T_{c2}$, then the system undergoes a phase transition to Branch (5,+,-).
We can determine the entropies in these two branches at this point,
\be
S_1(T=T_{c2})=26.7242 k T_{c2},\ S_5(T=T_{c2})=17.7638 k T_{c2}.
\ee
The change of the entropy at this point is
\be
\Delta S=S_5-S_1=-8.96039 k T_{c2}.
\ee
We can also find a point $T_2$ that the entropies in Branch (1,+,-) and (5,+,-) become equal. It is
\be
\mu T_2^2=-0.00636156
\ee
with the corresponding entropy
\be
S_1(T=T_2)=S_5(T=T_2)=21.1408 k T_2.
\ee
The Fig. \ref{STSpin4neg} shows the relation between $S$ and $T$ in different branches for a negative chemical potential.

 However,
we also notice that Branch (5,+,-) is unstable at the critical point $T_{c2}$, where the specific heat in Branch (5,+,-) is negative. Only after $T=T_{c2}'>T_{c2}$, the specific heat becomes positive.
We can determine the turning point $T=T_{c2}'$ by analyzing the extreme value of (\ref{Lmu}) in Branch (5,+,-). The extreme point is at \be
z=z_{c2}'=-0.1766
\ee
with the corresponding $\mathcal{L}$ and $T_{c2}'$ being
\be
\mathcal{L}=-\frac{0.4793}{\mu}=14.20 T_{c2}'^2,\ \ \mu T_{c2}'^2=-0.03376.
\ee
After that the system remains in Branch (5,+,-) to arbitrary high temperature.

There are two remarkable facts. The first is that the critical values are different for $\mu>0$ and $\mu<0$, a reflection of the charge conjugate violation found in \cite{truncated}. 
The second is that there exist thermodynamically unstable regions in phase diagram, in which the specific heat of the system is negative. For $\mu>0$, the thermodynamically unstable region is $T_{c1}<T<T_{c1}'$. For $\mu<0$, the thermodynamically unstable region is $T_{c2}<T<T_{c2}'$.


\section{Conclusion and Discussion}

In this paper, we explored the phase structure of the higher spin black holes in $AdS_3$. We found a general entropy formula via dimensional analysis and thermodynamics. For $SL(N,R)$ case, this formula depends on $(N-1)$ pairs of conjugate variables $\mathcal{L}_n,\alpha_n$. After some small modification, it can be used to  $Sp(2N,R), SO(2N+1,R), G_2$ and even  $hs(\lambda)$  gravity. We checked that it indeed led to  the correct branches for the spin 3 and the spin $\tilde{4}$ black holes. It is now clear that the  dimensional analysis is valid for all the branches and the holonomy conditions could be used to relate the conjugate variables. More explicitly, for $SL(N,R)$ gravity, the holonomy conditions can be used to reduce half of the variables. Since we have obtained the general formula of the entropy and can always solve the holonomy equations numerically, we may work out in principle the phase structure for arbitrary higher spin black holes. It would be interesting to explore the phase structure  of the higher spin black hole in the large N limit and thus check the Gaberdial-Gopakumar conjecture even in the finite temperature, non-perturbative region. It is also interesting to check the finite higher spin version of the conjecture by exploring the phase structure on both sides.

In this paper, however, we were not so ambitious. We restricted ourselves to the higher spin black holes which can be solved exactly. The motivation to restrict to the spin 3 and spin $\tilde{4}$ black holes  is two-fold. Firstly, we need to check the consistency of our method. Secondly,  it is better to have an exact solution capturing the quantitative behavior of the phase structure of the higher spin black holes. And this will be the cornerstone to the more complicated case. The study of the spin 3 and $\tilde{4}$ black holes shows that the $BTZ$ branch is indeed only dominated in the low temperature region. In the high temperature, the systems should transit to other branches, depending on the signature of the chemical potentials. The spin $\tilde{4}$ black hole is a little more complex than the spin 3 case, due to the charge conjugate asymmetry and the existence of unstable region. However, we cannot say that all higher spin black holes share the same feature since we only explore two simplest rank 2 cases. For another rank 2 $G_2$ case, following  the convention  in \cite{truncated}, the solutions of the holonomy equations are very like the spin $\tilde{4}$ black hole except that there are twelve branches now.  The parameter $z$ is defined as $z=\frac{\mathcal{W}_6}{\mathcal{L}^3}$, in terms of which the entropy functions are respectively
\bea
S_i&=&2\pi k\sqrt{\mathcal{L}}\cos\frac{1}{6}(\arcsin{\frac{1}{343} (143 - 36450 z)}-\arcsin{\frac{143}{343}}+2(i-1)\pi),\hs{3ex}i\leq 6 \nn\\
S_{i+6}&=&2\pi k\sqrt{\mathcal{L}}\cos\frac{1}{6}(\arcsin{\frac{1}{343} (143 - 36450 z)}+\arcsin{\frac{143}{343}}+(2i-1)\pi),\hs{3ex}i\leq 6 \nn
\eea
The phase structure should be more complicated but we expect that the basic picture is the same. We leave this theory for future study.

As the rank of the group increases, the branches of the higher spin black hole will grow exponentially. So it is important for us to know how many branches there are and how many phase transitions will occur as the temperature increases. Hopefully the problem can be solved since the partition function and the entropy formula are known. The only remaining problem is to solve the holonomy equations. An analytic solution may be unavailable, but a numerical simulation can be done to an arbitrary explicit level.

Another interesting problem we have ignored completely is the role of RG flow. As suggested in  \cite{Ammon:2011,David:2012}, the IR degrees of freedom are the degrees of freedom in the principal embedding while the UV degrees of freedom may be the degrees of freedom in the non-principal embedding. The IR CFT is deformed by an irrelevant higher spin operator hence flows to a different UV CFT. While from the UV CFT point of view, the UV CFT is deformed by a relevant operator hence flows to a different IR CFT. For the black holes with only one single higher spin hair, the RG flow analysis is relatively easy and could be done. We hope to return to this issue in the future.

\vspace*{10mm}
\noindent {\large{\bf Acknowledgments}}\\
The work was in part supported by NSFC Grant No. 10975005, ~11275010.

\end{document}